\documentclass[nolineno]{JFM-FLM_Au}

\usepackage{floatrow}
\usepackage{graphicx}
\usepackage{epstopdf, epsfig}
\usepackage{caption}      
\usepackage{subcaption}  
\usepackage{xcolor}
\usepackage{comment}
\newcommand{\comt}[1]{}
\usepackage{amsmath}

\lefttitle{Flapping Dynamics of a Compliant Membrane}
\righttitle{C.-Y. Zhang, A.-K. Gao and X.-J. Zhu}

\title{Flapping Dynamics of a Compliant Membrane in a uniform incoming flow}

\author{Chengyao Zhang\aff{1}, Ankang Gao\aff{2}  \and Xiaojue Zhu\aff{1}}

\affiliation{\aff{1}Max Planck Institute for Solar System Research,
G\"ottingen 37077, Germany
\aff{2}Department of Modern Mechanics, University of Science and Technology of China, Hefei, Anhui 230026, China}

\corresau{Xiaojue Zhu, \email{zhux@mps.mpg.de}}

\begin{document}
\maketitle

\begin{abstract}
Recent theoretical and experimental investigations have revealed that flapping compliant membrane wings can significantly enhance propulsive performance (e.g. Tzezana \& Breuer, 2019, J. Fluid Mech., 862, 871–888) and energy harvesting efficiency (e.g. Mathai et al., 2022, J. Fluid Mech., 942, R4) compared to rigid foils. Here, we numerically investigate the effects of the stretching coefficient (or aeroelastic number), $K_S$, the flapping frequency, $St_c$, and the pitching amplitude, $\theta_0$, on the propulsive performance of a compliant membrane undergoing combined heaving and pitching in uniform flow. Distinct optimal values of $K_S$ are identified that respectively maximize thrust and efficiency:  thrust can be increased by 200\%, and efficiency by 100\%, compared to the rigid case.  Interestingly, these optima do not occur at resonance but at frequency ratios (flapping to natural) below unity, and this ratio increases with flapping frequency. 
Using a force decomposition based on the second invariant of the velocity gradient tensor $Q$, which measures the relative strength between the rotation and deformation of fluid elements, we show that thrust primarily arises from 
$Q$-induced and body-acceleration forces. The concave membrane surface can trap the leading-edge vortex (LEV) from the previous half-stroke, generating detrimental $Q$-induced drag. However, moderate concave membrane deformation weakens this LEV and enhances body-acceleration-induced thrust. Thus, the optimal $K_S$ for maximum thrust occurs below resonance, balancing beneficial deformation against excessive drag.
Furthermore, by introducing the membrane's deformation into a tangential angle at the leading edge and substituting it into an existing scaling law developed for rigid plates, we obtain predictive estimates for the thrust and power coefficients of the membrane. The good agreement confirms the validity of this approach and offers insights for performance prediction.

\end{abstract}

\begin{keywords}
\end{keywords}


\section{Introduction}
\label{sec:Introduction}
Flapping wings or tails are frequently employed in the locomotion of aerial and aquatic animals \citep{wang2005dissecting,wu2011fish,liu2024vortices}. On the one hand, this is because biological systems do not support rotor-like spinning machines. On the other hand, the flapping wing has more aerodynamic or hydrodynamic advantages at moderate Reynolds numbers, such as high lift, low noise, and agility \citep{lighthill1969hydromechanics,eldredge2019leading,Jaworski2020Owl}. Over billions of years of evolution, animals have adapted to various environments and developed highly efficient locomotion capabilities. Studying the dynamic and biological mechanisms underlying these abilities serves as a rich source of inspiration for bio-inspired design. 

The rigid plate with prescribed pitching and heaving motion has been widely used in the modeling of the animal wing due to its simplicity and close relevance to most biological systems \citep{triantafyllou1991wake,anderson1998oscillating,dong2006wake,green2011unsteady,ayancik2019scaling,zhu2021nonlinear,chao2024tailbeat,jin2024enhancing}. For instance, \cite{buchholz2008wake} experimentally investigated a finite aspect ratio panel undergoing pitching motion around its leading edge in a uniform incoming flow and found that the panel's thrust coefficient increases monotonically with the Strouhal number, aspect ratio, and Reynolds number. Experimental studies by \citep{von2003flow,buchholz2006evolution} explored the effects of these parameters on the vortex structures behind the flapping plate, while \cite{quinn2014unsteady} examined the influence of a solid boundary on the performance of flapping wings undergoing pitching motion in a free-stream flow. Regarding the development of a generic scaling analysis for flapping foils, \cite{floryan2017scaling} presented scaling relations for the thrust coefficient, power coefficient, and propulsive efficiency of rigid foils undergoing either heaving or pitching motions, considering unsteady lift and added mass forces. They demonstrated that the mean thrust for heaving is entirely lift-based, while for pitching, it is solely due to added mass. However, the mean input power and propulsive efficiency depend on contributions from both mechanisms. These scaling laws were validated through water tunnel experiments on a nominally two-dimensional flapping plate in a free stream, as well as by biological data of swimming aquatic animals. In a related study, \cite{van2019scaling} experimentally investigated rigid foils undergoing combined heaving and pitching motions in a free stream, considering the effect of phase difference. Their findings revealed that the pitching motion should lag the heaving motion by about $30^\circ$ to maximize thrust and by about $90^\circ$ to maximize the propulsive efficiency. Expanding on the scaling law developed by \cite{floryan2017scaling} for individual heaving or pitching motions, \cite{van2019scaling} developed new scaling laws for combined heaving and pitching motions, considering the phase offset between them, which were validated by experimental data.

Inspired by the role of flexibility in aerial locomotion, particularly in structures such as flying animals' wings \citep{wootton1992functional, brodsky1994evolution}, researchers have studied passive or active deformation, in addition to passive flapping \citep{spagnolie2010surprising,zhang2010locomotion}, to gain deeper insights into how structural flexibility influences propulsive performance \citep{liao2003fish,shyy2010recent,zhu2014flow,zhu2014flexibility,gazzola2015gait,hoover2018swimming,smits2019undulatory,liu2024vortices}.
Experimental studies conducted by \cite{thiria2010wing} and \cite{ramananarivo2011rather} investigated the self-propulsion of flexible plates in the air. They found that flapping frequencies lower than the natural frequency of the flexible plate maximize the thrust coefficient and propulsive efficiency. Their experimental observations matched well with predictions from a nonlinear oscillator model, which incorporated cubic structural nonlinearities and both linear and quadratic damping terms. The modeling further revealed that non-resonant behavior emerges both in the presence of nonlinear damping at small flapping amplitudes and due to geometric saturation effects at large amplitudes. Therefore, resonance mechanisms cannot explain the observed performance optimum. Instead, they found that at the optimal frequency ratio, the instantaneous angle of attack and the trailing-edge deflection angle become equal at the moment of maximum flapping velocity, a configuration that is more favorable for generating aerodynamic pressure contributing to thrust. \cite{kang2011effects} numerically investigated the flapping dynamics of flexible plates. They found that the maximum propulsive force is achieved slightly lower than resonance. In contrast, the optimal propulsive efficiency occurs when the flapping frequency is approximately half the natural frequency. A non-dimensional tip-deformation parameter was proposed to characterize the structural response, and both the time-averaged force and efficiency were shown to scale with this parameter. \cite{alben2012dynamics} observed resonant-like peaks in the self-propulsive speed as the length and rigidity of the flexible plate were varied, both in numerical simulations and experiments. The inviscid model predicted that the self-propulsive speed scales with the length and rigidity, and these scalings were validated by experimental data. It was also found that the optimal propulsive efficiency can occur below, at, or above the resonant frequency \citep{ramananarivo2011rather,michelin2009resonance,masoud2010resonance}. All of these cases were observed in the experimental study \citep{dewey2013scaling} on flexible pitching panels in free-stream flow. Therefore, there is no consensus on whether resonance plays a definitive role in maximizing thrust and propulsive efficiency.

It is important to note that the above-mentioned investigations into flexible plates treat those plates as inextensible. In comparison, compliant membranes possess inherent stretchability, allowing them to adapt to changing flow conditions through passive deformation and elastic extension\citep{swartz1996mechanical,shyy1999flapping,song2008aeromechanics,cheney2015wrinkle,mavroyiakoumou2020large,mavroyiakoumou2022membrane,mathai2023shape}. Unlike rigid or merely inextensible plates, these stretchable membranes have been demonstrated to reduce flow separation and delay stall, and research on these aerodynamic performance improvements of membrane wings has been reviewed in  \cite{lian2003membrane} and \cite{tiomkin2021review}. Specifically, in theoretical investigations, \cite{waldman2017camber} presented a model that incorporates the Young–Laplace equation to capture the nonlinear deformation of a membrane at low angles of attack and employs thin-airfoil theory to approximate the aerodynamic pressure acting on the membrane. The predicted camber, lift, and vibrational frequency show good agreement with numerical simulations and experimental observations. Furthermore, \cite{tzezana2019thrust} developed a theoretical framework that combines thin airfoil theory with a membrane equation to model both steady and flapping membranes. As the flapping frequency increases, the membranes transition from thrust to drag near the resonant frequency, with this transition occurring earlier for more compliant membranes. Additionally, the wake experiences a transition from a reverse von K\'{a}rm\'{a}n vortex street to a traditional von K\'{a}rm\'{a}n vortex street. For highly compliant membranes, the wake transition frequency is predicted to be higher than the thrust–to-drag transition frequency. In the domain of numerical simulations and experiments, \cite{lauber2023rapid} employs a fully coupled fluid–structure interaction approach to investigate the effects of the Strouhal number and membrane compliance on the flight performance of bat-like flapping membrane wings, specifically designed to mimic the biomechanics of the handwing. They showed that a peak in both propulsive and lift efficiencies occurs near $St \approx 0.5$, which is significantly higher than the classical optimal Strouhal range observed in bird and fish locomotion. This shift is attributed to the specialized kinematics of bat flight. Moreover, they also found that while reducing membrane stiffness improves propulsive efficiency, excessive compliance induces flutter and significantly degrades performance, and introducing anisotropy into the material model helps delay the onset of flutter. In contrast, the numerical study by \citet{kumar2025computational} incorporates both the handwing and the armwing, enabling a more complete representation of bat wing kinematics. They employ a force partitioning method to analyze the aerodynamic forces and show that the dominant contribution arises from Q-induced forces, while the time-averaged contribution of kinematic forces is minimal. However, the time variation of the various partitions of the pressure-induced lift and drag forces reveals that the primary source of force oscillations is the kinematic component. This insight highlights the important role of unsteady inertial effects, particularly those associated with wing flutter. Similarly, \cite{gehrke2025highly} experimentally investigated the effects of the stretching stiffness coefficient and pitching amplitude on the lift and efficiency of a flapping membrane in a stationary fluid. In their setup, the membrane undergoes a prescribed horizontal heaving motion while maintaining a constant pitching angle throughout the flapping cycle. They identified distinct optimal values for the stretching stiffness coefficient and angle of attack, which maximize lift and efficiency, respectively. They observed different vortex patterns, and, unlike the rigid plate, it is no longer the leading-edge vortex that generates lift. Instead, the curvature of the membrane suppresses the leading-edge vortex. Vorticity then accumulates in a bound shear layer that covers the entire membrane, thereby enhancing lift. Beyond bio-inspired studies of flapping flight, compliant membranes have also been investigated for their potential in flow energy harvesting. For example, \citet{mathai2022fluid} experimentally studied a flexible membrane oscillating in a uniform flow and demonstrated that it could extract up to $160\%$ more power than a rigid plate. This enhancement was attributed to the increased lateral force resulting from the delayed shedding of the leading-edge vortex.

Despite the existing theoretical, numerical, and experimental works on stretchable membranes, fundamental physical questions remain unanswered: How does membrane stretchability at different levels alter the strength of vortex structures formed on the membrane and impact its dynamic performance? And how can we capture these coupled fluid–structure dynamics in a unified scaling framework that predicts thrust and power? To address these questions, we perform fully coupled fluid–structure interaction simulations of a two-dimensional membrane undergoing simultaneous heaving and pitching in a uniform flow, while systematically varying the membrane’s stretching stiffness, Strouhal number, and pitching amplitude. By analyzing the evolution of vortices and membrane dynamics, we uncover the physical mechanisms through which propulsion performance is enhanced and modify scaling laws that capture the dominant contributions to thrust and power, with good agreement across a wide range of parameters.

The remainder of this paper is organized as follows. The physical problem and  
mathematical formulation are presented in § \ref{Sec:problem&formulation}. The numerical method is described in § \ref{Sec:method}. Detailed results are discussed in § \ref{sec:Results}, and concluding remarks are addressed in § \ref{sec:Conclusion}.

\section{Computaional model}\label{sec:Computational model}
\subsection{Physical problem and mathematical formulation}\label{Sec:problem&formulation}
\begin{figure}[htbp]
  \centerline{\includegraphics[width=0.8\textwidth]{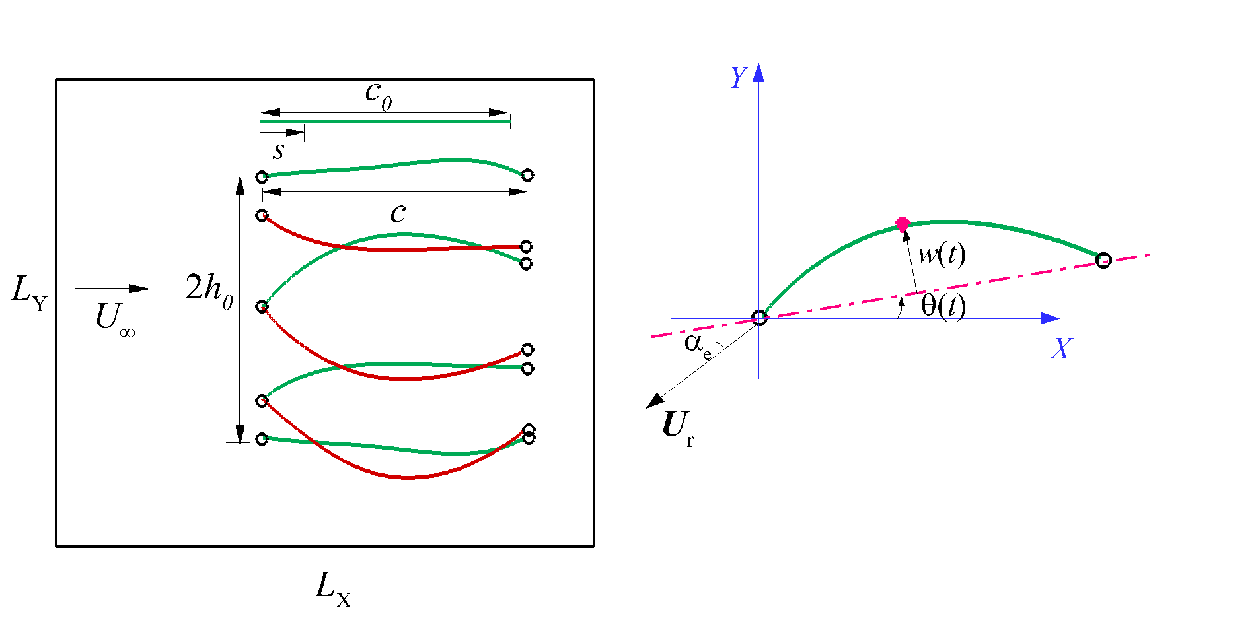}}
  \caption{Schematic diagram of a membrane flapping in a uniform incoming flow. Here, $h_0$ denotes the heaving amplitude, and $\theta(t)$ represents the instantaneous pitching angle.}
\label{Fig1_scheme}
\end{figure}

A two-dimensional model of a stretchable membrane flapping in uniform flow is considered. As shown in figure~\ref{Fig1_scheme}, a compliant membrane with an original length $c_0$ in the flat and stress-free state is pre-stretched to a length $c$ and immersed in a viscous fluid. The computational
domain is an $L_X\times L_Y$  rectangle, where $L_X$ and $L_Y$ are the domain sizes in the $x-$ and $y-$ directions, respectively.
To investigate this system of fluid-membrane interaction, the incompressible Navier-Stokes equations are solved to simulate the flow,
\begin{eqnarray}
\label{Eqn:NS} \rho \frac{\partial \boldsymbol{v}}{\partial
t}+\rho \boldsymbol{v}\cdot\nabla\boldsymbol{v}&=&-\nabla
p+\mu\nabla^2\boldsymbol{v}+\boldsymbol{f},\\
\label{Eqn:Continuity} \nabla\cdot\boldsymbol{v}&=&0,
\end{eqnarray}
where $\boldsymbol{v}$ is the velocity, $p$ the pressure, $\rho$ the density of
the fluid, $\mu$ the dynamic viscosity, and $\boldsymbol{f}$ the body force term.

As shown in figure~\ref{Fig1_scheme}, the leading edge of the membrane undergoes a heaving motion, while the trailing edge experiences a combination of heaving and pitching motions about the leading edge. The green curves represent the membrane during the downstroke, whereas the red curves represent the membrane during the upstroke. The heaving and pitching motions are respectively described by the following equations:
\begin{eqnarray}
\label{Eqn:motion0}
h(t)=h_0 \cos(2\pi f t),\quad \theta(t)=\theta_0 \cos(2\pi f t+\phi), 
\end{eqnarray}
where the phase difference $\phi$ between the heaving and pitching motions is set to $-90^\circ$. In figure~\ref{Fig1_scheme}, $w(t)$ is the instantaneous maximal deflection relative to the undeformed state, with positive values representing upward deflections, and negative values representing downward deflections. $\alpha_{e}$ is the effective attack of angle, defined as
\begin{eqnarray}
\label{Eqn:AOA}
\alpha_{e}(t)=-\theta(t)-\tan^{-1}\left(\frac{\dot{h}(t)}{U_\infty}\right).
\end{eqnarray}
The deformation and motion of the membrane are described by the structural equation~\citep{connell2007flapping,hua2013locomotion} 
\begin{eqnarray}
\label{Eqn:plate}
\rho_s \frac{\partial^2 \boldsymbol{X}}{\partial t^2} = \frac{\partial}{\partial s}\left[ Eh\left(1-1/\left| \frac{\partial \boldsymbol{X}}{\partial s}\right|\right)\frac{\partial \boldsymbol{X}}{\partial s}-\frac{\partial}{\partial s}\left(EI\frac{\partial^2\boldsymbol{X}}{\partial s^2}\right)\right] + \boldsymbol{F}_{s},
\end{eqnarray}
where $s$ is the Lagrangian coordinate along the membrane in its flat, stress-free state, with $s\in[0,c_0]$ and $\boldsymbol{X}(s,t)$ is the Eulerian position vector for a Lagrangian point $s$ at time $t$. The parameters $Eh$ and $EI$ represent the stretching and the bending stiffnesses, respectively. A pre-stretch is introduced through the following boundary conditions:
\begin{align*}
\label{Eqn:BC1}
\boldsymbol{X}(0,t) &=(0,h(t)),\\
\boldsymbol{X}(c_0,t) &=(c \cos\theta(t),h(t)-c \sin\theta(t)).
\end{align*}
Here, $c$ is the length after pre-stretching, with a pre-stretch ratio of $\lambda_0=c/c_0$. Additionally, the simply supported boundary condition is applied at both ends as follows:
\begin{eqnarray*}
\label{Eqn:BC2}
\frac{\partial^2 \boldsymbol{X}}{\partial s^2}=(0,0),
\frac{\partial^3 \boldsymbol{X}}{\partial s^3}=(0,0),
\end{eqnarray*}
The above equations are non-dimensionalized by the following characteristic scales: the chord length $c$, the fluid density $\rho$, and the incoming flow velocity $U_\infty$. The dimensionless governing parameters are listed as follows: the Reynolds number $Re=\rho U_\infty c/\mu$, the stretching coefficient $K_S=Eh/\rho U_\infty^{2}c$ (The parameter is also referred to in the literature as the aeroelastic number \citep{tzezana2019thrust}), the bending coefficient $K_B=EI/\rho U_\infty^{2} {c}^{3}$, the mass ratio $M=\rho_s/\rho c$, and the Strouhal number $St_c=fc/U_\infty$. Moreover, the instantaneous thrust coefficient (\(\widetilde{C_T}\)), time-averaged thrust coefficient (\(C_T\)), time-averaged power coefficient (\(C_P\)) and propulsive efficiency (\(\eta\))~\citep{van2019scaling} are defined as:
\begin{equation}
\begin{gathered}
\widetilde{C_T}=\frac{-\int_0^c F_x(s,t) \, ds}{\frac{1}{2}\rho U_{\infty}^2 c},\;
C_T=\frac{1}{T_f}\int_t^{t+T_f}\widetilde{C_T} \, dt, \\
C_P=\frac{1}{T_f}\frac{-\int_t^{t+T_f}\int_0^c \boldsymbol{F}_{s}(s,t)\cdot \partial_t \boldsymbol{X} \, ds \,dt}{\frac{1}{2}\rho U_{\infty}^3 c},\;
\eta=\frac{C_T}{C_P}
\end{gathered}
\label{Eqn:CTCP}
\end{equation}

\subsection{Numerical method and validation}\label{Sec:method}

The lattice Boltzmann method (LBM) \citep{chen1998lattice} is employed to solve the governing equations of fluid flow. The corresponding lattice Boltzmann equation with a body force model \citep{guo2002discrete} is given by:
\begin{equation}
\label{Eqn:LBE}
f_i(\boldsymbol{x} + \boldsymbol{e}_i \delta t, t + \delta t) - f_i(\boldsymbol{x}, t) = - \frac{1}{\tau}(f_i(\boldsymbol{x}, t) - f_i^{eq}(\boldsymbol{x}, t)) + \delta t \boldsymbol{F}_i,
\end{equation}
where $f_i(\boldsymbol{x}, t)$ is the distribution function for particles with velocity $\boldsymbol{e}_i$ at position $\boldsymbol{x}$ and time t, $\tau$ is the non-dimensional relaxation time, and $\delta t$ is the time increment. The $D2Q9$ velocity model is applied here. The equilibrium distribution function $f_i^{eq}$ and the force term $\boldsymbol{F}_i$ \citep{chen1998lattice,guo2002discrete} are defined as

\begin{equation}
\label{Eqn:feq}
f_i^{eq}= w_i \rho [1+\frac{ \boldsymbol{e}_i \cdot \boldsymbol{v} }{ c_s^2 } + \frac{\boldsymbol{v}\boldsymbol{v}:(\boldsymbol{e}_i \boldsymbol{e}_i - c_s^2\boldsymbol{I})}{2c_s^2}],
\end{equation}
\begin{equation}
\label{Eqn:Fi}
\boldsymbol{F}_i = (1-\frac{1}{2\tau})w_i[\frac{\boldsymbol{e}_i - \boldsymbol{v}}{c_s^2} + \frac{\boldsymbol{e}_i\cdot\boldsymbol{v}}{c_s^4}\boldsymbol{e}_i ]\cdot \boldsymbol{f},
\end{equation}
where $w_i$ is the weighting factor and $c_s$ is the lattice sound speed. The density $\rho$ and velocity $\boldsymbol{v}$ can be calculated by the distribution functions
\begin{equation}
\label{Eqn:rho}
\rho = \sum_{i} f_i,
\end{equation}
\begin{equation}
\rho\boldsymbol{v} = \sum_{i} \boldsymbol{e}_{i}f_{i} + \frac{1}{2}\boldsymbol{f}\delta t.
\end{equation}

The structure equation (\ref{Eqn:plate}) is solved by the finite element method with the co-rotational scheme \citep{doyle2013nonlinear}. The immersed boundary method (IBM) couples the fluid-structure interaction~\citep{peskin2002immersed}. The Lagrangian force exerted on the membrane by the fluid $\boldsymbol{F}_{s}=(F_x,Fy)$ in Equation~(\ref{Eqn:plate}) is calculated using the penalty scheme \citep{goldstein1993modeling}
\begin{equation}
\label{Eqn:penalty}
\boldsymbol{F}_{s}(s,t)=\alpha\int_{0}^{t}[\boldsymbol{V}_f(s,t^{'})-
\boldsymbol{V}_s(s,t^{'})]\mathrm{d} t^{'}
+\beta[\boldsymbol{V}_f(s,t)-\boldsymbol{V}_s(s,t)],
\end{equation}
where $\alpha$ and $\beta$ are penalty parameters, and the corresponding values are selected based on the previous studies
\citep{hua2014dynamics,peng2018hydrodynamic,zhang2020effect},
$\boldsymbol{V}_s=\frac{\partial \boldsymbol{X}}{\partial t}$ is the velocity of the Lagrangian point of the plate, and $\boldsymbol{V}_f$ is the fluid velocity at the position $\boldsymbol{X}$ obtained by interpolation
\begin{equation}
\label{Eqn:Vf}
\boldsymbol{V}_f(s,t)=\int \boldsymbol{v}(\boldsymbol{x},t) \delta(\boldsymbol{x}-\boldsymbol{X}(s,t))\mathrm{d} \boldsymbol{x},
\end{equation}
where $\delta(\boldsymbol{x}-\boldsymbol{X})$ is the smoothed Dirac delta function. The body force term $\boldsymbol{f}$ in Equation~(\ref{Eqn:NS}) is used as an interaction force between the fluid and the immersed boundary to enforce the no-slip velocity boundary condition. The body force $\boldsymbol{f}$ on the Eulerian points can be obtained from the Lagrangian force $\boldsymbol{F}_s$ using the Dirac delta function, i.e.,
\begin{eqnarray}
\label{Eqn:f}
\boldsymbol{f}(\boldsymbol{x},t)=-\int \boldsymbol{F}_{s}(\boldsymbol{x},t) \delta(\boldsymbol{x}-\boldsymbol{X}(s,t))\mathrm{d} s.
\end{eqnarray}

\begin{figure}[htbp]
  \centering
  \begin{subfigure}[b]{\textwidth}
      \centering
      \includegraphics[width=\textwidth]{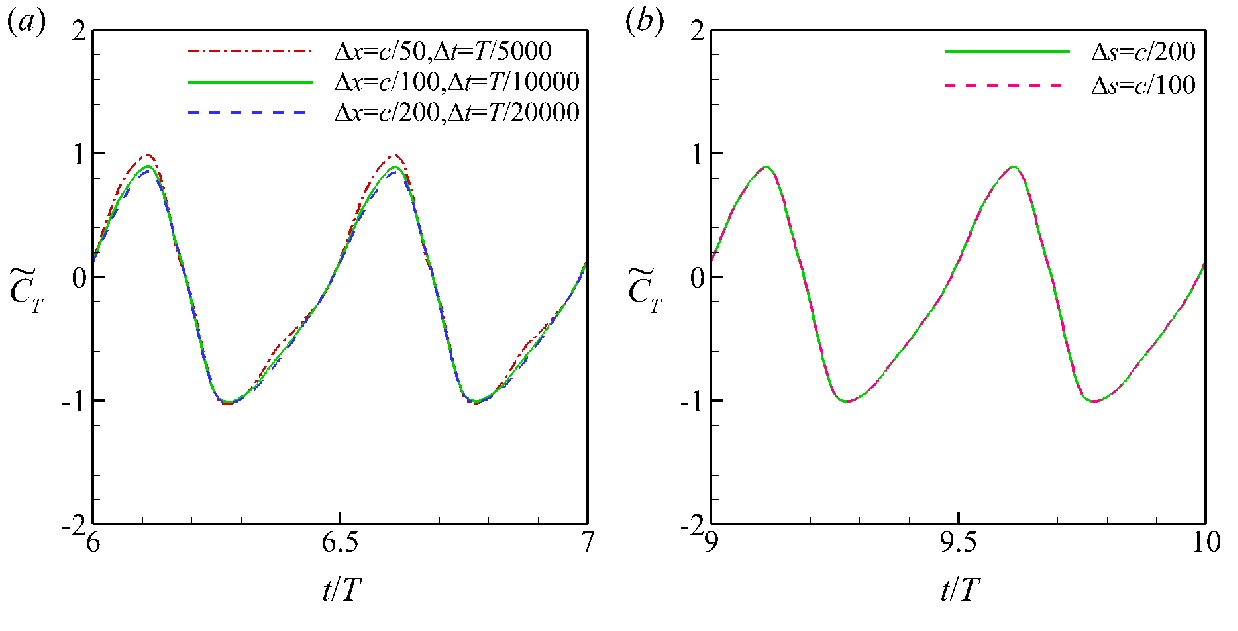}
  \end{subfigure}
  \begin{subfigure}[b]{\textwidth}
      \centering
      \includegraphics[width=\textwidth]{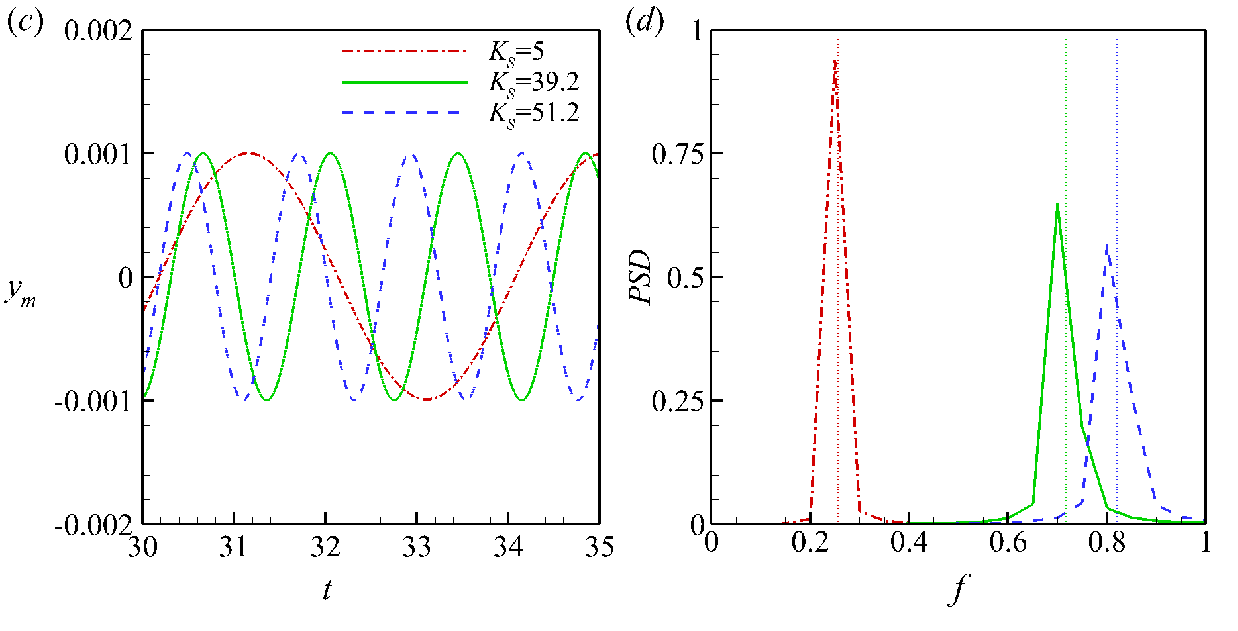}
  \end{subfigure}
  \caption{Time variations of $\widetilde{C_T}$ for cases with $Re=200$, $St_c=0.25$, $K_S=10$, $\theta_0=15^\circ$ and $M=1.0$, simulated under ($a$) three different combinations of grid spacings and time steps, and ($b$) two discrete mesh sizes for the membrane. ($c$) Time evolution of the dimensionless transverse displacement $y_m$ of the midpoint of a membrane with fixed simply supported boundaries at both ends, undergoing free vibrations triggered small perturbations for three different values of $K_S$. ($d$) Power spectral density (PSD) of $y_m$ corresponding to the three $K_S$ cases shown in ($c$), and the dotted lines indicate the analytical values of the first-order natural frequencies for each case.}
  \label{Fig2_convergence}
\end{figure}

A uniform velocity $U_\infty$ is applied at the inlet boundary and the two side boundaries of the fluid computational domain. A Neumann boundary condition ($\partial \boldsymbol{U}/\partial x=0$) is specified at the outlet boundary. The computational domain for fluid flow is chosen to be $40c\times 20c$ in the $x$ and $y$ directions to eliminate boundary effects. A convergence study for grid spacing and time step is conducted for cases with $Re=200$, $St_c=0.2$, $K_S=30$, $\theta_0=15^\circ$, and $M=1.0$. figure~\ref{Fig2_convergence}(a) shows the time-dependent thrust coefficient $\widetilde{C_T}$ under three different grid spacings $\Delta x$ and time steps $\Delta t$. It can be seen that $\Delta x=c/100$ and $\Delta t=T/10000$ are sufficiently small to yield accurate results, where $T$ is the flapping period. Therefore, $\Delta x=c/100$ and $\Delta t=T/10000$ are adopted for all the present simulations. Since the membrane in this study can undergo stretching, two different structure grid spacings, $\Delta s$, are tested. Figure~\ref{Fig2_convergence}(b) presents the time-dependent thrust coefficient for a grid with the same size as the fluid grid, $\Delta s=c/100$, and a finer grid, $\Delta s=c/200$, used to discretize the solid structure. Although the different values of $\Delta s$ do not produce any changes, a finer solid grid with $\Delta s=c/200$ is used to better capture the membrane deformation.

The numerical strategy used here has been successfully applied to flexible plates without pre-stretching in a
wide range of flows\citep{hua2014dynamics,peng2018hydrodynamic,liu2022scaling, peng2022scaling,xiong2024numerical}. Figure~\ref{Fig2_convergence}($c$) and ($d$) validate the FEM solver for simulating the free vibrations of a membrane with $M=1.0$, $K_B=10^{-4}$, and a pre-stretching ratio of $\lambda_0=1.05$ in a vacuum. The membrane is subject to small perturbations and has fixed simply supported boundary conditions at both ends. The validation is performed by comparing the simulation results with the analytical solution for the natural frequency. Figure~\ref{Fig2_convergence}($c$) presents the time-dependent transverse displacement (\(y_m\)) of the membrane’s midpoint for \(K_S = 5\), \(K_S = 39.2\), and \(K_S = 51.2\). The power spectral density ($PSD$) analysis of \(y_m\), shown in figure~\ref{Fig2_convergence}(d), reveals that the dimensionless dominant frequencies ($f$) are $0.25$, $0.699999988$, and $0.800000012$ for \(K_S = 5\), \(K_S = 39.2\), and \(K_S = 51.2\), respectively. The analytical expression for the dimensionless \(n\)-th natural frequency is given by
\begin{eqnarray}
\label{Eqn:fn}
f_n = \frac{n}{2} \sqrt{\frac{K_S (\lambda_0 - 1)}{M/\lambda_0}}.
\end{eqnarray}
For the first-order natural frequencies ($f_1$), the analytical values for \(K_S = 5\), \(K_S = 39.2\), and \(K_S = 51.2\) are $0.2562$, $0.7173$, and $0.819756$, respectively. The relative error, given by $\frac{\lvert f-f_1 \rvert}{f_1}$, remains within $2.5\%$, indicating strong agreement between the simulation results and the analytical solution.

\section{Results and discussion}\label{sec:Results}
In our study, the Reynolds number, mass ratio, heaving amplitude, pre-stretch ratio, and bending stiffness coefficient are set to $Re=200$, $M=1.0$, $h_0/c=0.5$, $\lambda_0=1.05$, and $K_B=10^{-4}$, respectively. The values of $M$ and $\lambda_0$ are selected based on previous studies\citep{jaworski2015thrust,waldman2017camber,tzezana2019thrust}. The effects of the stretching stiffness coefficient ($1<K_S<\infty$, where $K_S=\infty$ represents a rigid plate), the Strouhal number ($0.2\leq St_c \leq 0.4$), and the pitching amplitude ($10^\circ\leq\theta_0\leq20^\circ$) on the propulsive performance of the compliant membrane are investigated.

\subsection{Propulsive performance}
\begin{figure}[htbp]
  \centerline{\includegraphics[width=\textwidth]{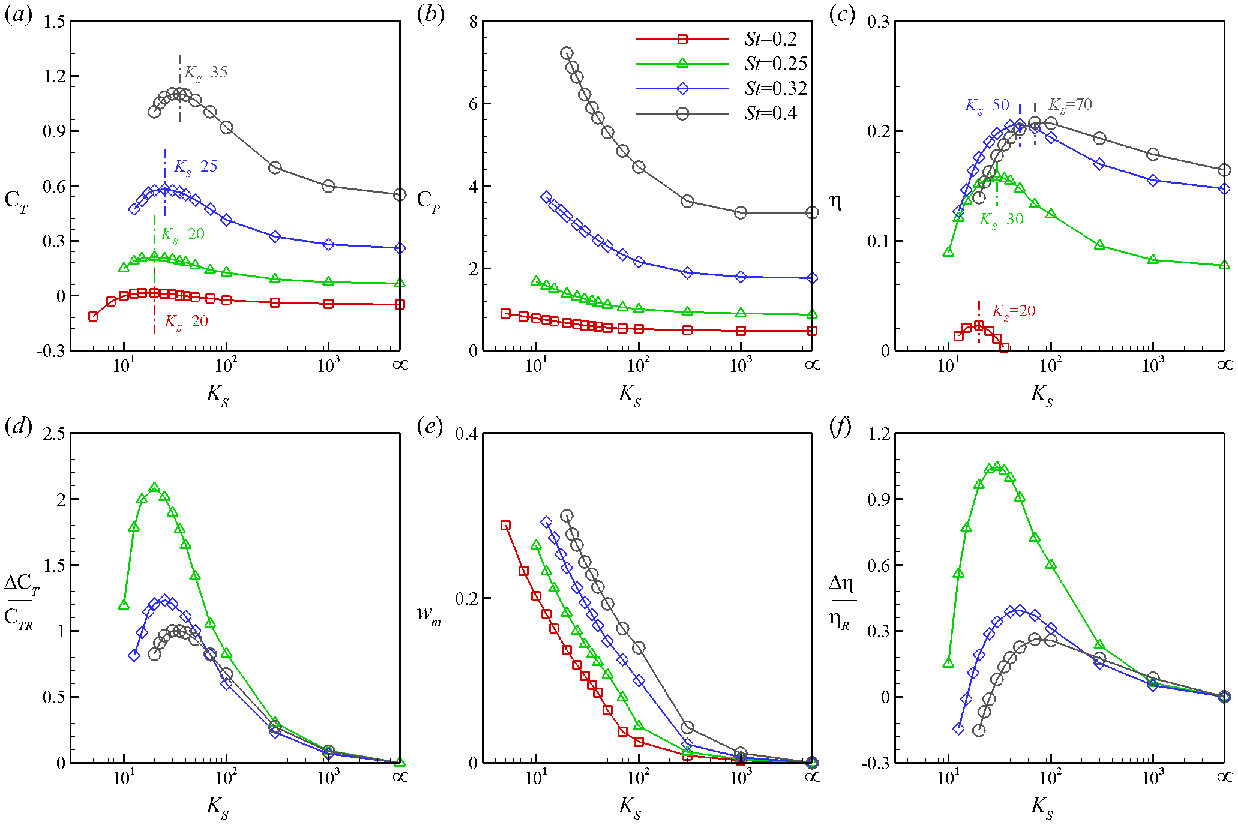}}
  \caption{(a) Time-averaged thrust coefficient $C_T$, (b) time-averaged power coefficient $C_P$, (c) propulsive efficiency $\eta$, (d) thrust increment ratio $\frac{\Delta C_T}{C_{TR}}$, (e) maximum deflection over a cycle $w_m$ as functions of $K_S$ for $\theta_0=15^{\circ}$, and (f) propulsive efficiency increment ratio $\frac{\Delta \eta}{\eta_R}$.}
\label{Fig3_S}
\end{figure}

Figure~\ref{Fig3_S} shows the propulsive performance metrics as functions of the stretching coefficient $K_S$ for a representative pitching amplitude of $\theta_0=15^{\circ}$. As shown in figure~\ref{Fig3_S}(a), for all considered Strouhal numbers $St_c$, the time-averaged thrust coefficient $C_T$ increases with $K_S$ initially and then decreases, indicating the existence of an optimal $K_S$ ($K_S^*$) that maximizes thrust. Furthermore, as $St_c$ increases from $0.2$ to $0.4$, $K_S^*$ also increases from $K_S^*=20$ to $K_S^*=30$.To quantify the propulsive performance enhancement of the membrane compared to the rigid plate, relative increment ratios are introduced. Figure~\ref{Fig3_S}(d) shows the thrust increment ratio $\Delta C_T/C_{TR}=(C_T-C_{TR})/C_{TR}$ as a function of $K_S$, where $C_{TR}$ is the time-averaged thrust coefficient of the rigid plate. It should be noted that the definition of $\Delta C_T/C_{TR}$ is valid only when $C_{TR} > 0$. For $St_c=0.2$, the rigid plate experiences drag, whereas the membrane, due to the effect of stretching, generates thrust. For larger $St_c$, stretching enables the membrane to achieve a thrust enhancement that is $200\%$ of the rigid plate's thrust, and $\Delta C_T/C_{TR}$ decreases as $St_c$ increases for each $K_S$.

Figure~\ref{Fig3_S}(b) shows that the time-averaged power coefficient 
$C_P$ decreases monotonically as $K_S$ increases for each $St_c$. Combined with the trend of the time-averaged thrust coefficient with respect to $K_S$, this indicates that, in general, higher thrust is achieved at the expense of increased input power. In addition, according to the definition of the power coefficient, since the motion in the streamwise direction is highly constrained (i.e., $\partial_t X(s,t)$ is very small), the input power mainly originates from the contributions of the force and velocity in the transverse ($y$) direction. Therefore, we introduce the maximum deflection relative to the undeformed state of the membrane across the chord at each instant during a flapping period, denoted as $w(t)$, while $w_m$ represents the maximum absolute value of $w$ over a full cycle. Figure~\ref{Fig3_S}(e) shows that $w_m$ decreases monotonically as $K_S$ increases. This reduction in $w_m$ results in a smaller lateral displacement of the membrane during a flapping period, thereby reducing the lateral velocity. Consequently, the decrease in $w_m$ partially explains the monotonic decline in the power coefficient with increasing $K_S$.

Since $C_T$ varies non-monotonically with $K_S$, while $C_P$ varies monotonically with $K_S$, the propulsive efficiency $\eta=C_T/C_P$ also varies non-monotonically with $K_S$, as shown in figure~\ref{Fig3_S}(c). Similar to $C_T$, there exists $K_S^*$ that maximizes the propulsive efficiency. Furthermore, as $St_c$ increases from $0.2$ to $0.4$, $K_S^*$ increases from $20$ to $70$. Figure~\ref{Fig3_S}($f$) shows the propulsive efficiency increment ratio $\Delta \eta/\eta_R=(\eta-\eta_R)/\eta_R$ as a function of $K_S$, where $\eta_R$ is the propulsive efficiency coefficient of the rigid plate. The propulsive efficiency increment ratio follows a similar trend to the thrust increment ratio as $St_c$ and $K_S$ change, and stretching can increase the membrane’s propulsive efficiency by up to $100\%$ compared to the rigid plate.

To analyze the dynamic response of the membrane, the frequency ratio $f^*$, defined as the ratio of the flapping frequency to the first-order natural frequency of the membrane in a uniform incoming flow (the latter is deduced in  \cite{tzezana2019thrust}), is introduced as:
\begin{eqnarray}
\label{Eqn:freqr}
f^*=\frac{f}{f_1}=2 St_c \sqrt{\frac{M/\lambda_0 + M_A}{K_S (\lambda_0-1)}},
\end{eqnarray}
where $M_A$ is the added mass coefficient and $M_A = 0.5$ is obtained in \cite{tzezana2019thrust}.

\begin{figure}[htbp]
  \centerline{\includegraphics[width=\textwidth]{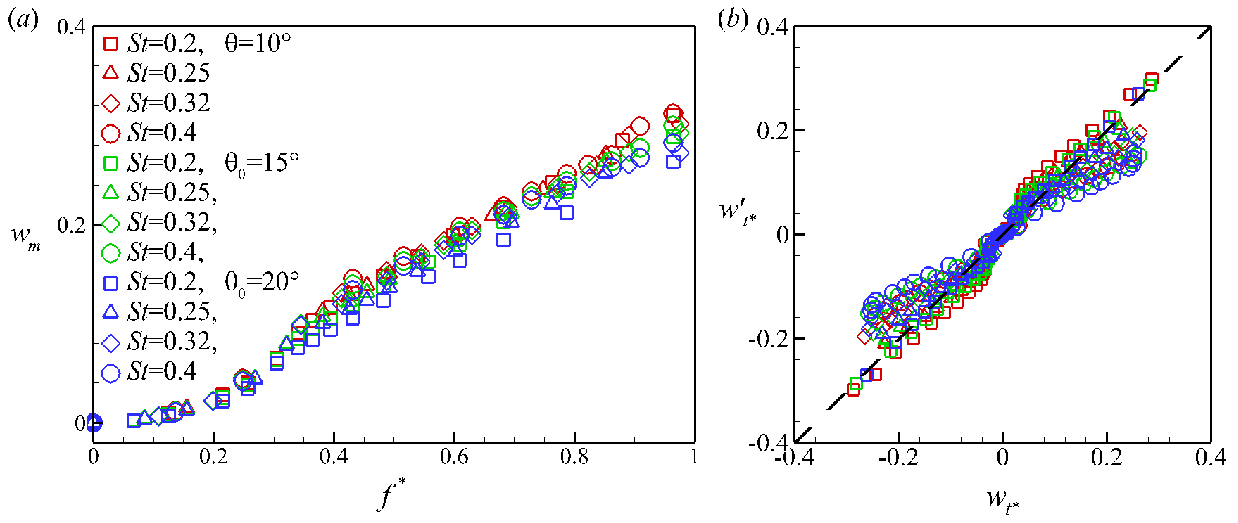}}
  \caption{(a) Variation of $w_m$ as a function of $f^*$ for four different values of $St_c$ (represented by different symbol shapes), with $\theta_0 = 10^{\circ}$ (red symbols), $\theta_0 = 15^{\circ}$ (green symbols), and $\theta_0 = 20^{\circ}$ (blue symbols). (b) Predicted $w^{\prime}_{t^*}$ versus numerically simulated $w_{t^*}$ at $t=\frac{T}{4}$ and $t=\frac{3T}{4}$ for all $\theta_0$ and $St_c$.
 }
\label{Fig4_wm_f}
\end{figure}

Figure~\ref{Fig4_wm_f}(a) shows the variation of $w_m$ with the frequency ratio $f^*$ for three different pitching amplitudes ($\theta_0=10^\circ$, $15^\circ$, and $20^\circ$). $f^*$ nearly collapses $w_m$ from different $St_c$ and $\theta_0$ onto a single curve, and $w_m$ increases monotonically with $f^*$ before reaching the resonance. For the instantaneous maximal deflection $w(t)$, it can be approximated as a steady-state response for simplicity and derived based on the Young-Laplace equation\citep{waldman2017camber,mathai2022fluid}:
\begin{eqnarray}
\label{Eqn:YLE0}
\kappa+\frac{p}{\mathcal{T}}=0,
\end{eqnarray}
where $\kappa$ is the curvature, $p$ is the pressure difference across the membrane, and $\mathcal{T}$ is the tension. Assuming the deformation of the membrane forms an arc, the following can be derived from the geometric relationship: $\kappa=8w/(1+4w^2)$, $\lambda=(2\lambda_0/\kappa)\sin^{-1}(\kappa/2)$ and $\mathcal{T}=Eh(\lambda-1)$. The pressure across the membrane is calculated by the thin-airfoil theory \citep{waldman2017camber},
\begin{eqnarray}
\label{Eqn:pressure}
p=\pi \rho U_\infty^2\sin(\alpha_{e}+\frac{1}{2}\sin^{-1}(\kappa/2)),
\end{eqnarray}
where $\alpha_{e}$ is the instantaneous effective attack of angle. By substituting the expressions for $\alpha_{e}$, pressure $p$, and curvature $\kappa$ into Young-Laplace equation (\ref{Eqn:YLE0}), and nondimensionalizing using the membrane's fixed-end length $c$ as the length scale and $\rho U_\infty^2$ as the pressure scale, the resulting nondimensional equation is expressed as:
\begin{eqnarray}
\label{Eqn:YLE}
\frac{\pi\sin(\alpha_{e}+\frac{1}{2}\sin^{-1}(\tilde{\kappa}(\tilde{w}/2)))}{(\lambda(\tilde{w})-1)\tilde{k}(\tilde{w})}=K_S,
\end{eqnarray}
where $\tilde{}$ represents the nondimensionalized quantities.

Figure~\ref{Fig4_wm_f}(b) presents the instantaneous maximal deflection at two instants, $t^* = \frac{T}{4}$ and $t^* = \frac{3T}{4}$, where $T$ is the flapping period. The instantaneous maximal deflections, obtained from equation~(\ref{Eqn:YLE}) and numerical simulations, are denoted as $w_{t^*}^\prime$ and $w_{t^*}$, respectively. The results show good agreement between the two approaches. The discrepancy between $w_{t^*}^\prime$ and $w_{t^*}$ increases with $St_c$, and can be explained as follows. $w_{t^*}^\prime$ is derived based on steady-state theory, which assumes that the deformation is a symmetric circular arc. However, a larger $St_c$ corresponds to a higher flapping frequency, where dynamic effects become more pronounced, and the deformation deviates from a symmetric circular arc, as shown in figures~\ref{Fig9_S25_wQ} and~\ref{Fig10_S12.5_wQ}. These dynamic deviations lead to a greater divergence of the $w_{t^*}^\prime$–$w_{t^*}$ comparison from the $x = y$ line.

\begin{figure}[htbp]
  \centering
  \begin{subfigure}[b]{\textwidth}
      \centering
      \includegraphics[width=\textwidth]{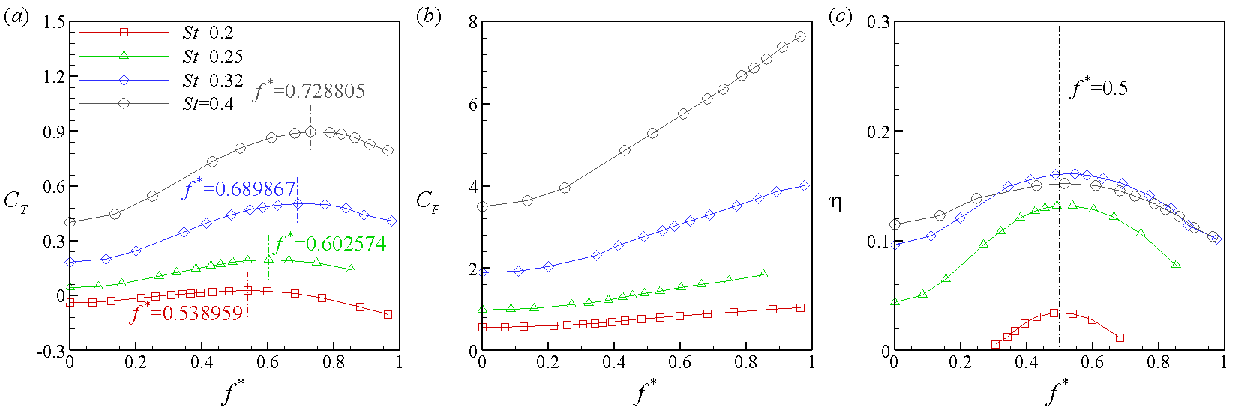}
  \end{subfigure}
  \begin{subfigure}[b]{\textwidth}
      \centering
      \includegraphics[width=\textwidth]{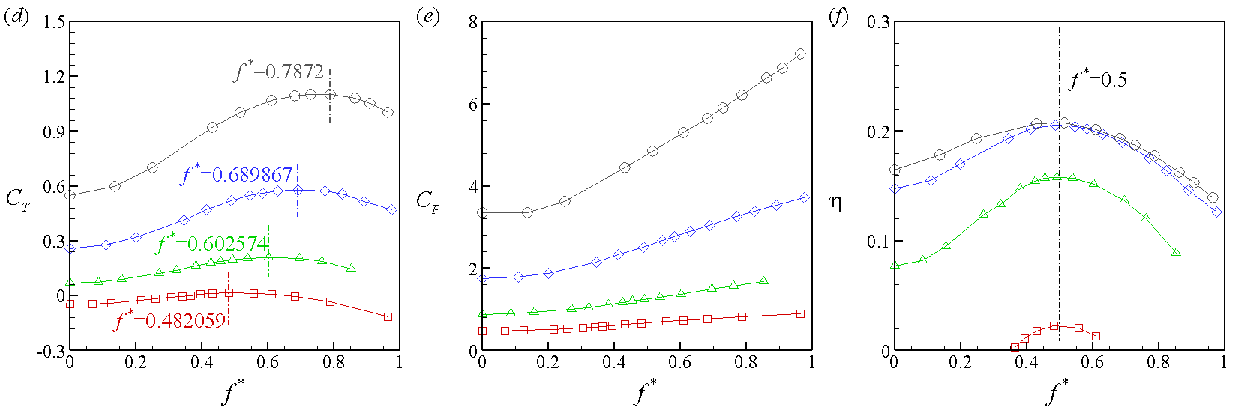}
  \end{subfigure}
    \begin{subfigure}[b]{\textwidth}
      \centering
      \includegraphics[width=\textwidth]{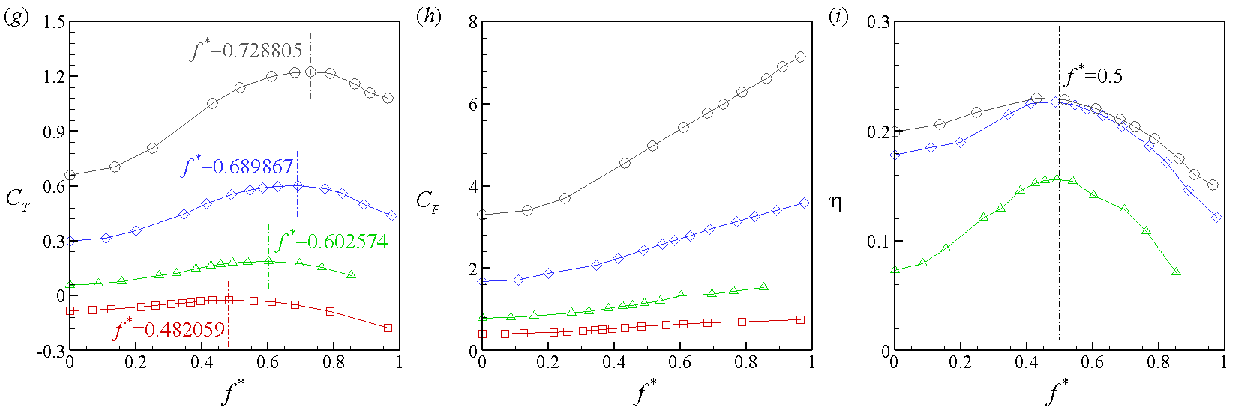}
  \end{subfigure}
  \caption{($a$,$d$,$g$) The thrust coefficient $C_T$, ($b$,$e$,$h$) power coefficient $C_P$, and ($c$,$f$,$i$) propulsive efficiency $\eta$ as functions of $f^*$ for ($a$-$c$) $\theta_0=10^{\circ}$, ($d$–$f$) $\theta_0=15^{\circ}$, and ($g$–$i$) $\theta_0=20^{\circ}$, respectively.}
  \label{Fig5_statistics}
\end{figure}

Figure~\ref{Fig5_statistics} presents the propulsive performance metrics ($C_T$, $C_P$, and $\eta$) as functions of the frequency ratio ($f^*$) for three different pitching amplitudes ($\theta_0=10^\circ$, $15^\circ$, and $20^\circ$). It can be observed that for each combination of $\theta_0$ and $St_c$, the thrust coefficient ($C_T$) first increases and then decreases with increasing $f^*$, exhibiting an optimal frequency ratio ($f^*_{opt}$) that maximizes $C_T$. The values of $f^*_{opt}$ are explicitly marked in figure~\ref{Fig5_statistics}($a$,$d$,$g$). This trend is consistent with the theoretical results of \cite{tzezana2019thrust}, which indicate that as $f^*$ increases, the thrust initially rises and then transitions to drag near resonance. Additionally, for each $\theta_0$, as $St_c$ increases, $f^*_{opt}$ also increases. However, for a given $St_c$, $\theta_0$ has little influence on the value of $f^*_{opt}$. Figure~\ref{Fig5_statistics}($b$,$e$,$h$) shows that the time-averaged power coefficient ($C_P$) increases monotonically with $f^*$ for each combination of $St_c$ and $\theta_0$. Furthermore, for a given $\theta_0$, $C_P$ increases monotonically with $St_c$ at a fixed $f^*$ within the considered range of $St_c$.

Similar to the thrust coefficient, the propulsive efficiency ($\eta$) first increases and then decreases with $f^*$ for each combination of $St_c$ and $\theta_0$, exhibiting an optimal frequency ratio ($f^*_{opt}$) that maximizes $\eta$, as shown in figure~\ref{Fig5_statistics}($c$,$f$,$i$). However, unlike $C_T$, the optimal $f^*$ for $\eta$ remains nearly fixed at $0.5$, corresponding to a flapping frequency that is half of the membrane's resonance frequency. This behavior arises because, for each $\theta_0$, while $f^*_{opt}$ for $C_T$ increases with $St_c$, the monotonic increase in $C_P$ with $f^*$ becomes more pronounced as $St_c$ increases. Consequently, based on the definition of propulsive efficiency, $\eta = C_T / C_P$, the optimal $f^*$ for $\eta$ remains unchanged with varying $St_c$.

\subsection{Unsteady dynamics}

To gain a clearer understanding of the thrust experienced by the membrane and to identify how the membrane enhances thrust compared to a rigid flapping plate, we employ a force decomposition method based on a weighted integral of the second invariant of the velocity gradient tensor. This method ensures that the decomposition results are Galilean invariant~\citep{gao2019note}. The details can be found in \cite{gao2019passing,gao2023three} and \cite{menon2021quantitative} are briefly summarized as follows.

The pressure Poission equation in incompressible flows is derived by taking the divergence of the Navier–Stokes equations, given by 
\begin{eqnarray}
\label{Eqn:PPE}
\nabla^2 p = -\rho \nabla \cdot (\boldsymbol{v} \cdot \nabla \boldsymbol{v}) = 2 \rho Q,
\end{eqnarray}
where \( Q = -\frac{1}{2} \nabla \boldsymbol{v} : (\nabla \boldsymbol{v})^T = \frac{1}{2} ( \|\boldsymbol{\Omega}\|^2 - \|\boldsymbol{D}\|^2 ) \) is the second invariant of the velocity gradient tensor \( \nabla \boldsymbol{v} \), with \( \boldsymbol{\Omega} \) and \( \boldsymbol{D} \) denoting the antisymmetric and symmetric parts of \( \nabla \boldsymbol{v} \), representing the rotation-rate and strain-rate tensors, respectively. Here, \( \| \cdot \| \) denotes the Frobenius norm. The Q-criterion, which is based on this quantity, is widely used for vortex identification.

By applying Green's identity, the following expression can then be derived:
\begin{equation}
\label{Eqn:px0}
\int_{\partial B} p \hat{\boldsymbol{n}} \cdot \hat{\boldsymbol{e}}_1 \, dS 
= \int_{V_{f\infty}} 2\rho Q \phi_1 \, dV 
- \int_{\partial B} \phi_1 \hat{\boldsymbol{n}} \cdot \nabla p \, dS - \int_{\Sigma} \left( \phi_1 \boldsymbol{n} \cdot \nabla p - p \boldsymbol{n} \cdot \nabla \phi_1 \right) dS,
\end{equation} 
where $\hat{\boldsymbol{e}}_i$ denotes the unit vector in the $i$-th coordinate direction ($i = 1, 2, 3$, corresponding to the $x$-, $y$-, and $z$-directions, respectively), and $\hat{\boldsymbol{n}}$ and $\boldsymbol{n}$ are the unit normal vectors pointing inward on the body surface and outward on the control surface, respectively. The auxiliary potentials $\phi_i$, introduced by \citet{quartapelle1983}, are harmonic functions that satisfy $\nabla^2 \phi_i = 0$ in the fluid domain, $\hat{\boldsymbol{n}} \cdot \nabla \phi_i = - \hat{\boldsymbol{n}} \cdot \hat{\boldsymbol{e}}_i$ on the body surface, and $\phi_i = 0$ in the far field at infinity. Physically, each $\phi_i$ characterizes the sensitivity of the hydrodynamic force in the $i$-th Cartesian direction with respect to local displacement of the body boundary, and depends solely on the instantaneous geometry of the body.

Since the surface integral over \(\Sigma\) vanishes as the control volume approaches the entire fluid domain \(V_{f\infty}\)~\citep{gao2019passing}, and the Neumann boundary condition for pressure on the body surface is given by
\[
-\hat{\boldsymbol{n}} \cdot \nabla p = \rho\, \hat{\boldsymbol{n}} \cdot \boldsymbol{a} - \mu\, \hat{\boldsymbol{n}} \cdot \nabla^2 \boldsymbol{v},
\]
where \(\boldsymbol{a} = D\boldsymbol{v}/D t\) denotes the local acceleration of the fluid, the \(x\)-component of the pressure force can be expressed as 

\begin{equation}
\label{Eqn:px}
\int_{\partial B} p \hat{\boldsymbol{n}} \cdot \hat{\boldsymbol{e}}_1 \, dS 
= \int_{V_{f\infty}} 2\rho Q \phi_1 \, dV 
+ \int_{\partial B} \rho \phi_1\, \hat{\boldsymbol{n}} \cdot \boldsymbol{a} \, dS 
- \int_{\partial B} \mu\, \phi_1\, \hat{\boldsymbol{n}} \cdot \nabla^2 \boldsymbol{v} \, dS.
\end{equation}

Therefore, the total force in the \(x\)-direction can be expressed as
\begin{equation}
\label{Eqn:Fx}
F_x = \int_{V_{f\infty}} 2\rho Q \phi_1 \, dV
+ \int_{\partial B} \rho \phi_1 \hat{\boldsymbol{n}} \cdot \boldsymbol{a} \, dS
+ \int_{\partial B} -\mu \phi_1 \hat{\boldsymbol{n}} \cdot \nabla^2 \boldsymbol{v} \, dS + \int_{\partial B} -2\mu\, (\boldsymbol{D} \cdot \hat{\boldsymbol{n}}) \cdot \hat{\boldsymbol{e}}_1
 \, dS.
\end{equation}

The four terms on the right-hand side of Equation~\eqref{Eqn:Fx} are denoted, respectively, by \(F_{px,Q}\), \(F_{px,a}\), \(F_{px,vis}\), and \(F_{x,f}\). $F_{px,Q}$ represents the contribution to the force from the source term in the pressure Poisson equation, and is referred to as the Q-induced force. A positive value of \(Q\) corresponds to a vortical structure, which generates an attractive pressure force on the body, whereas a negative value of \(Q\) corresponds to a strain-dominated region, producing a repulsive pressure force on the body. The second term, \(F_{p_x,a}\), arises from the acceleration of the plate. The third term, \(F_{p_x,\mathrm{vis}}\), represents the viscous contribution in the Neumann-type pressure boundary condition on the body surface. The fourth term, \(F_{x,f}\), corresponds to the frictional force. 

The instantaneous thrust coefficients corresponding to each of these four forces are denoted as 
\(\widetilde{C}_{T,Q}\), \(\widetilde{C}_{T,a}\), \(\widetilde{C}_{T,vis}\), and \(\widetilde{C}_{T,f}\), respectively, and are calculated as follows: 
\begin{equation}
\begin{gathered}
\widetilde{C}_{T,Q}=\frac{-F_{px,Q}}{\frac{1}{2}\rho U_{\infty}^2 c},
\widetilde{C}_{T,a}=\frac{-F_{px,a}}{\frac{1}{2}\rho U_{\infty}^2 c}, 
\widetilde{C}_{T,vis}=\frac{-F_{px,vis}}{\frac{1}{2}\rho U_{\infty}^2 c},
\widetilde{C}_{T,f}=\frac{-F_{x,f}}{\frac{1}{2}\rho U_{\infty}^2 c},
\end{gathered}
\label{Eqn:CT}
\end{equation}
and the time-averaged thrust coefficients are denoted by \(C_{T,Q}\), \(C_{T,a}\), \(C_{T,vis}\), and \(C_{T,f}\), respectively. 

\begin{figure}[htbp]
  \centerline{\includegraphics[width=\textwidth]{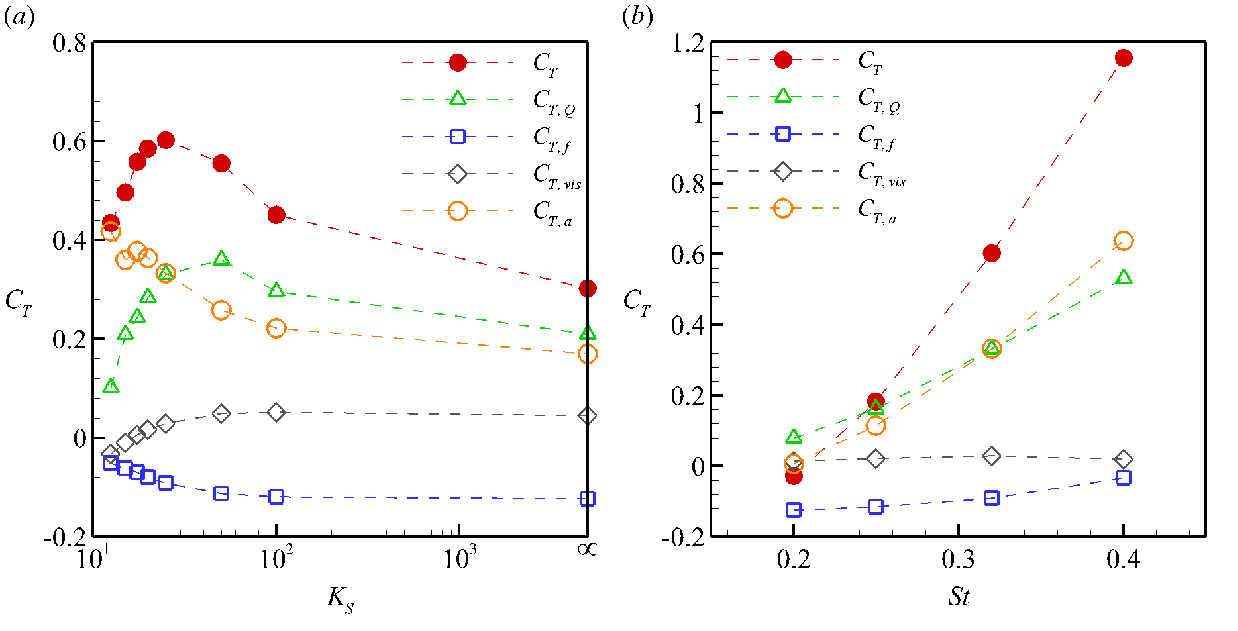}}
  \caption{Variation of time-averaged thrust coefficients: (a) with stiffness ratio $K_S$ at fixed Strouhal number $St_c = 0.32$, and (b) with Strouhal number $St_c$ at fixed stiffness ratio $K_S = 25$.
 }
\label{Fig6_avgCT}
\end{figure}

To isolate the effect of stretching on the propulsion performance of the membrane, figure~\ref{Fig6_avgCT}($a$) presents the variation of the time-averaged components of the thrust coefficient and their total with $K_S$ at a fixed pitching amplitude of $\theta_0 = 20^\circ$ and Strouhal number $St_c = 0.32$. The results indicate that the Q-induced force and the body-acceleration force are the primary contributors to thrust generation. The trend of the time-averaged Q-induced thrust coefficient $C_{T,Q}$ with increasing $K_S$ resembles that of the time-averaged total thrust coefficient $C_T$, it first increases and then decreases. The body-acceleration component $C_{T,a}$ decreases monotonically as $K_S$ increases, suggesting that greater stretchability leads to a higher contribution from body acceleration to the overall thrust. The frictional force component $C_{T,f}$ acts as a source of drag, and its magnitude increases with the increasing $K_S$, indicating that membrane deformation can reduce the friction drag. In contrast, the viscous pressure force contribution $C_{T,\mathrm{vis}}$ increases with $K_S$, but it remains small in magnitude throughout and can be neglected. Therefore, the non-monotonic behavior of the $C_T$ with increasing $K_S$ is caused by the non-monotonic response of $C_{T,Q}$. 
 
To isolate the effect of flapping frequency on the propulsion performance of the membrane, figure~\ref{Fig6_avgCT}($b$) presents the variation of the time-averaged thrust coefficient with $St_c$ at a fixed pitching amplitude of $\theta_0 = 20^\circ$ and stretching coefficient $K_S = 25$. The Q-induced force and the body-acceleration force are the primary contributors to thrust generation. A significant increase in $C_{T,Q}$ and $C_{T,a}$ is observed with increasing flapping frequency. The component $C_{T,f}$ contributes to drag, but its magnitude decreases with increasing frequency, thereby reducing the drag. The viscous pressure force component, $C_{T,\text{vis}}$, also decreases with frequency and remains negligibly small across the studied range. Hence, the overall thrust coefficient $C_T$ increases monotonically with increasing $St_c$. Moreover, the Q-induced force dominates the thrust at low frequencies, whereas the body-acceleration force grows faster with the increasing $St_c$ and surpasses the $C_{T,Q}$ at $St_c=0.32$. Together with the dependence of $C_T$, $C_{T,Q}$, and $C_{T,a}$ on $K_S$ as shown in figure~\ref{Fig6_avgCT}($a$), the competition between $C_{T,Q}$ and $C_{T,a}$ at different $St_c$ leads to a shift in the optimal $K_S$ that maximizes $C_T$ across varying $St_c$.

\begin{figure}[htbp]
  \centerline{\includegraphics[width=\textwidth]{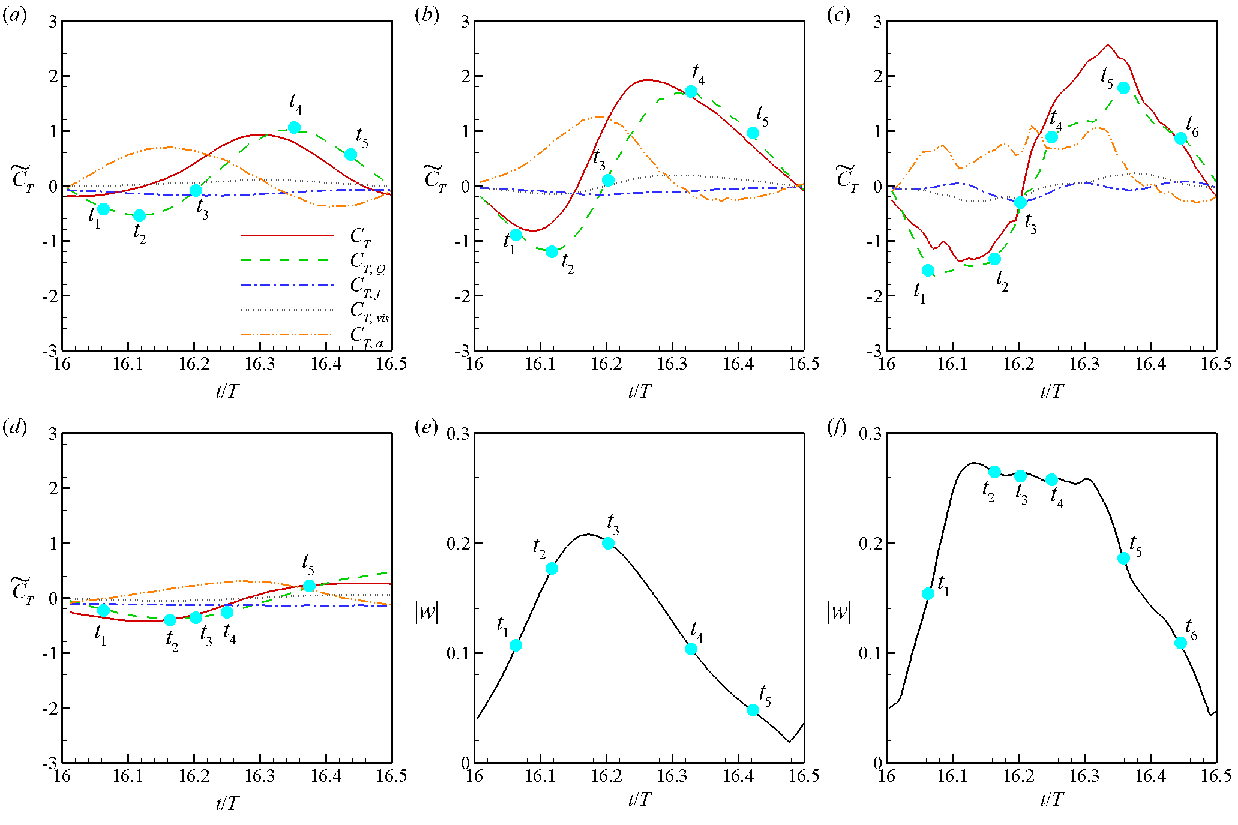}}
  \caption{Time variation of thrust coefficient during downstroke for (a) rigid plate, (b) $K_S = 25$, (c) $K_S = 12.5$ at $St_c = 0.32$, $\theta_0 = 20^\circ$ and (d) $K_S = 25$, $St_c = 0.2$, $\theta_0 = 20^\circ$. (e, f) Time variation of the absolute value of maximum deflection for the cases in (b) and (c), respectively.
}
\label{Fig7_CT_t}
\end{figure}

Figure~\ref{Fig7_CT_t}($a$–$c$) show the time-dependent force decomposition results corresponding to the cases with different $K_S$ in figure~\ref{Fig6_avgCT}($a$), during the downstroke phase. It can be seen that the amplitudes of $\widetilde{C}_{T,f}$ and $\widetilde{C}_{T,vis}$ are very small. Therefore, we focus on $C_{T,Q}$ and $C_{T,a}$ in the following analysis.  

For the rigid plate, as shown in figure~\ref{Fig7_CT_t}($a$), the plate undergoes acceleration followed by deceleration during the downstroke. As a result, the body-acceleration force contributes positively to thrust during the earlier portion of the downstroke and becomes a drag force later on. However, the magnitude of this drag contribution is relatively small. Therefore, the body-acceleration force remains the primary contributor to thrust generation over the entire flapping cycle. Compared with the rigid plate, the membrane with a stretching stiffness of $K_S = 25$ exhibits a significantly higher maximum $\widetilde{C}_{T,a}$, as shown in figure~\ref{Fig7_CT_t}($b$). This enhancement is likely due to the membrane stretching, which induces acceleration of a larger volume of surrounding fluid, resulting in a stronger reactive force and thus enhanced thrust. The time evolution of the maximum deflection over the entire membrane is shown in figure~\ref{Fig7_CT_t}($e$), and its peak occurs approximately at the same time as the peak in $\widetilde{C}_{T,a}$, indicating a strong correlation between membrane deformation and thrust enhancement. When the membrane is more compliant, as in the case of $K_S = 12.5$ shown in figure~\ref{Fig7_CT_t}($c$), $\widetilde{C}_{T,a}$ maintains a relatively high plateau over a prolonged period. This plateau coincides with the time interval during which the membrane exhibits a sustained large deflection as shown in figure~\ref{Fig7_CT_t}($f$). These observations suggest that membrane stretching can effectively enhance the body-acceleration component of thrust.

\begin{figure}[htbp]
  \centerline{\includegraphics[width=\textwidth]{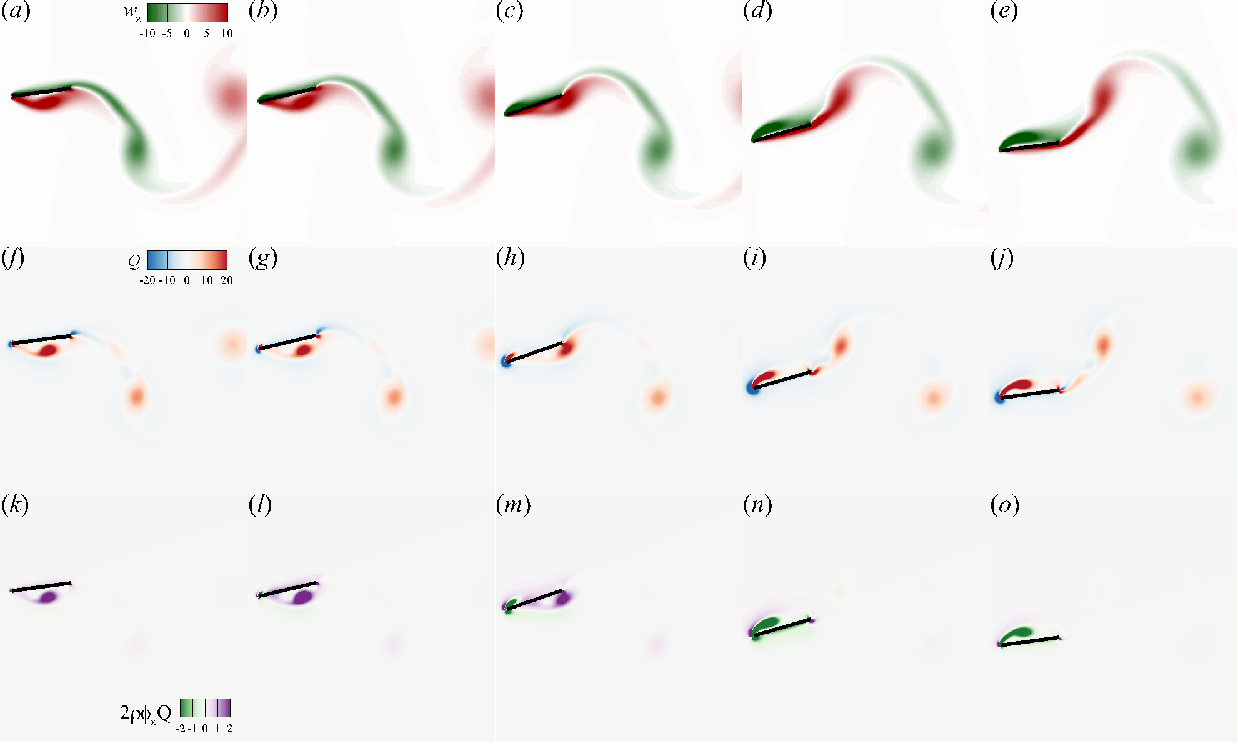}}
  \caption{Contours of (a-e) vorticity, (f-j) $Q$-criterion, and (k-o) $2\rho \phi_x Q$ for the rigid plate at five time instants: (a,f,k)$t_1$, (b,g,l)$t_2$,(c,h,m)$t_3$, (d,i,n)$t_4$, and (e,j,o)$t_5$ marked in figure~\ref{Fig7_CT_t}(a).}
\label{Fig8_rigid_wQ}
\end{figure}

Figures~\ref{Fig8_rigid_wQ}($a-e$) present the vorticity contours at five representative time instants $t_1$-$t_5$ (as marked in figure~\ref{Fig7_CT_t}) for the rigid plate. During this period, a leading-edge vortex (LEV) gradually forms and strengthens on the upper surface. Meanwhile, the vortex formed on the lower surface during the previous half-cycle convects downstream and detaches from the plate. Further insight is provided by the $Q$ contours in figures~\ref{Fig8_rigid_wQ}($f-j$), where a region with negative $Q$ values appears just upstream of the leading edge. This negative $Q$ region corresponds to areas where the strain rate dominates over rotation. The distribution of $2\rho \phi_x Q$ shown in figures~\ref{Fig8_rigid_wQ}($k-o$) highlights the contributions of these flow features to the Q-induced force. On the upper surface, the developing LEV induces suction in the negative $x$-direction, generating thrust that increases as the vortex grows. On the lower surface, the downstream-moving vortex induces suction in the positive $x$-direction, resulting in drag. Additionally, the negative $Q$ region contributes to a repulsive pressure effect: generating drag on the upper surface (positive $x$-direction) and thrust on the lower surface (negative $x$-direction), and these effects are relatively weak for the rigid plate. Overall, at the early stage of the downstroke (figures~\ref{Fig8_rigid_wQ}($k-m$)), LEV remains weak, while the suction induced by the vortex on the lower surface dominates, resulting in a net drag force as shown in figure~\ref{Fig7_CT_t}($a$) for $C_{T,Q}$. In the later stages of the downstroke (figures~\ref{Fig8_rigid_wQ}($n$-$o$)), the LEV strengthens and the vortex on the lower surface sheds, causing the Q-induced force to shift toward thrust.

\begin{figure}[htbp]
  \centerline{\includegraphics[width=\textwidth]{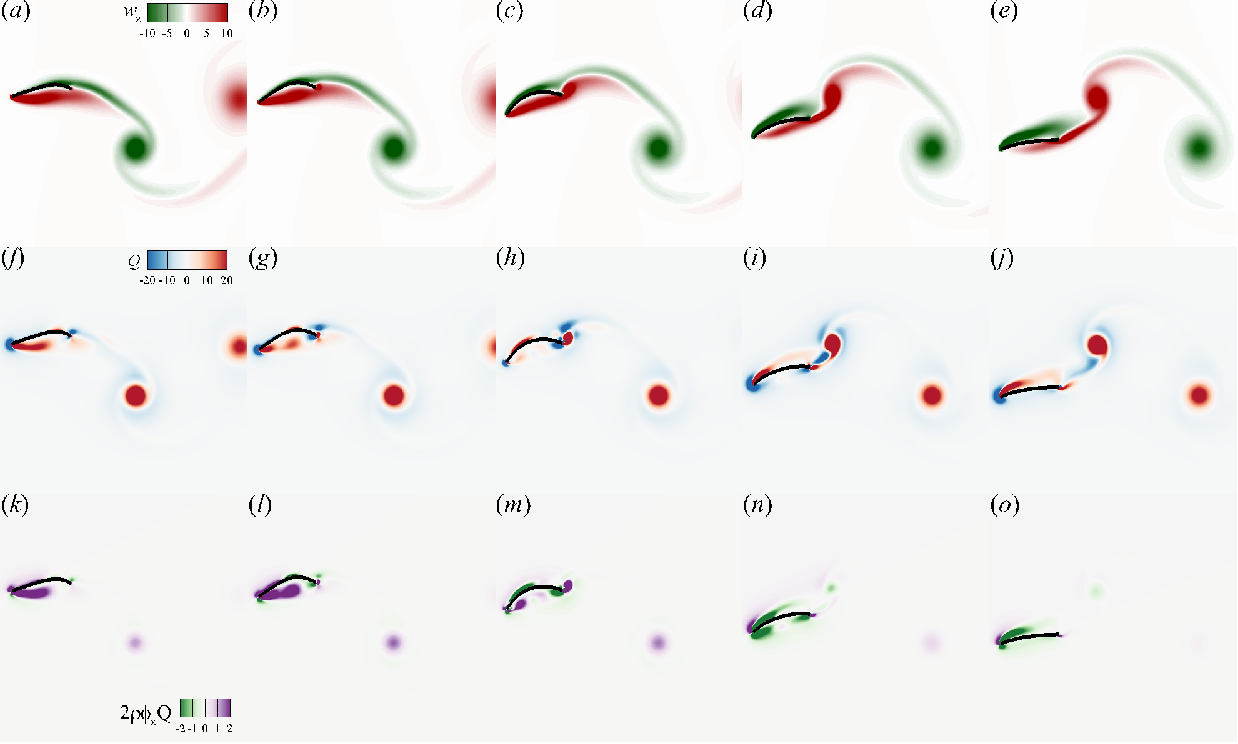}}
  \caption{Contours of (a-e) vorticity, (f-j) $Q$-criterion, and (k-o) $2\rho \phi_x Q$ for the membrane with $K_S=25$ at five time instants: (a,f,k)$t_1$, (b,g,l)$t_2$,(c,h,m)$t_3$, (d,i,n)$t_4$, and (e,j,o)$t_5$ marked in figure~\ref{Fig7_CT_t}.}
\label{Fig9_S25_wQ}
\end{figure}

Unlike the rigid plate, the membrane with $K_S$ exhibits an attached vorticity layer on its upper surface during the downward stroke, as shown in figures~\ref{Fig9_S25_wQ}($a$–$e$). On the other hand, the shear layer on the lower surface generated during the upward stroke keeps growing before being cut off by the trailing edge, forming a concentrated trailing-edge vortex. The leading-edge vortex beneath the lower surface is trapped (figures~\ref{Fig9_S25_wQ}($f$-$h$)), rather than being convected downstream as in the rigid plate case, inducing a vortex suction force in the positive $x$-direction and thereby generating a larger $Q$-induced drag force than that in the rigid plate during the early stage of the downward stroke, as illustrated in figures~\ref{Fig9_S25_wQ}($k$--$m$) and figure~\ref{Fig7_CT_t}($b$). During the later stage of the downward stroke, the upper-surface vorticity layer intensifies, and the resulting vortex suction force generates thrust in the negative $x$-direction, as shown in figures~\ref{Fig9_S25_wQ}($m$-$o$). Meanwhile, the negative-$Q$ region in front of the leading edge induces a thrust force in the negative $x$-direction on the lower surface, and a drag force in the positive $x$-direction on the upper surface, similar to the case of the rigid plate. However, the local curvature of the leading edge makes the repulsive pressure associated with this negative-$Q$ region appear to be stronger than that in the rigid plate case, as shown in figure~\ref{Fig9_S25_wQ}($n$). As a result, as shown in figure~\ref{Fig7_CT_t}($b$), the membrane experiences a net thrust force during the later stage, with the Q-induced thrust magnitudes exceeding that observed in the rigid plate case. Overall, the time-averaged $Q$-induced thrust coefficient $C_{T,Q}$ shows an increase compared to that of the rigid plate, as shown in figure~\ref{Fig6_avgCT}(a).

\begin{figure}[htbp]
  \centerline{\includegraphics[width=\textwidth]{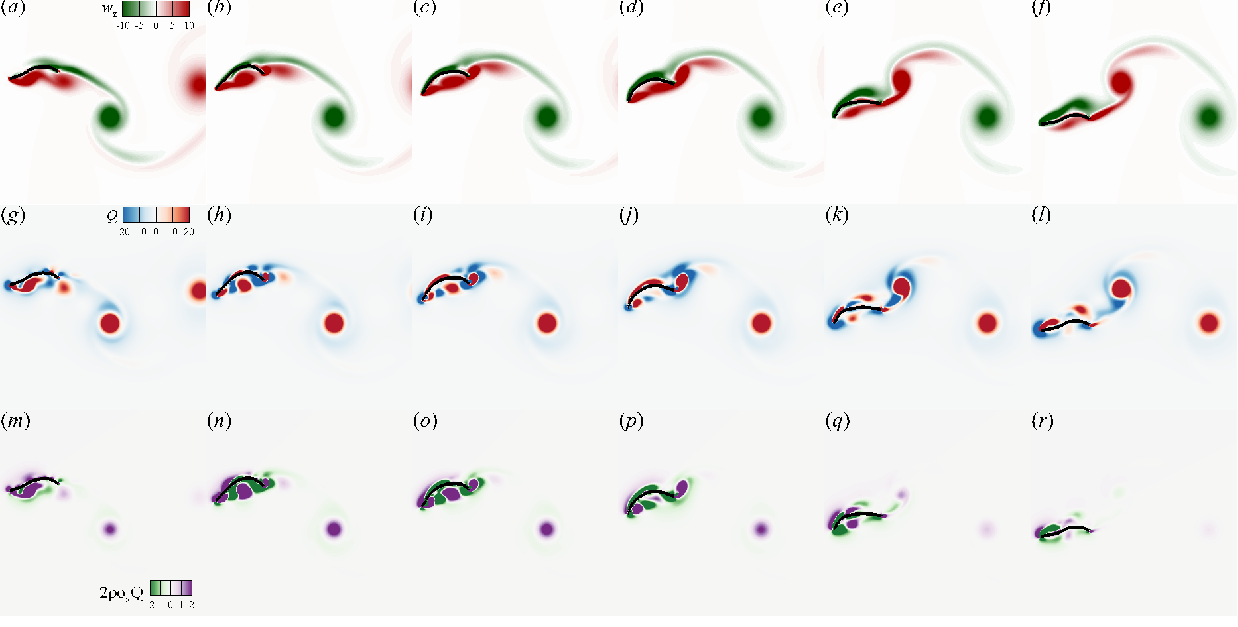}}
  \caption{Contours of (a-f) vorticity, (g-l) $Q$-criterion, and (m-r) $2\rho \phi_x Q$ for the membrane with $K_S=12.5$ at five time instants: (a,g,m)$t_1$, (b,h,n)$t_2$,(c,i,o)$t_3$, (d,j,p)$t_4$, (e,k,q)$t_5$, and (f,l,r)$t_6$ marked in figure~\ref{Fig7_CT_t}.}
\label{Fig10_S12.5_wQ}
\end{figure}

As $K_S$ further decreases, the increased upward bulging deformation of the membrane traps the leading-edge vortex beneath the lower surface and strengthens it into a more concentrated and rotational structure, as shown in figures~\ref{Fig10_S12.5_wQ}($a$–$c$) and ($g$–$i$) for $K_S=12.5$. The resulting vortex suction force contributes to the Q-induced drag (figures~\ref{Fig10_S12.5_wQ}($m$–$o$)) and is maintained at a higher level for a longer duration during the early stage of the downstroke, compared to the rigid plate and the $K_S=25$ case, as shown in figure~\ref{Fig7_CT_t}(c). On the upper surface, the vorticity layer strengthens as the membrane deformation increases (figures~\ref{Fig10_S12.5_wQ}($a$–$d$)). During the rebound from its peak deformation, a counter-sign vorticity layer develops between the membrane and the overlying primary layer (figures~\ref{Fig10_S12.5_wQ}($d$–$f$)). As a result, the overlying primary vorticity layer above the membrane becomes distorted and weakened, and is eventually reduced to a small residual structure near the leading edge (figures~\ref{Fig10_S12.5_wQ}($j$–$l$)). This degradation of the vorticity layer on the upper surface reduces the suction contribution to the Q-induced thrust, as evidenced by the distribution of $2\rho\phi_{x}Q$ shown in Figures~\ref{Fig10_S12.5_wQ}($p$–$r$). However, the presence of a negative-$Q$ region ahead of the leading edge generates additional thrust on the lower surface near the leading edge, partially compensating for the loss of suction on the upper surface. As a result, the peak value of the total Q-induced thrust does not decrease compared to the $K_S=25$ case in the later stage of the downstroke, as shown in Figure~\ref{Fig7_CT_t}($c$). Therefore, the capture and intensification of the leading-edge vortex—generated during the previous half-stroke beneath the lower surface—which enhances the Q-induced drag, leads to a significant reduction in the time-averaged value of $C_{T,Q}$ compared to the $K_S=25$ case.

\begin{figure}[htbp]
  \centerline{\includegraphics[width=\textwidth]{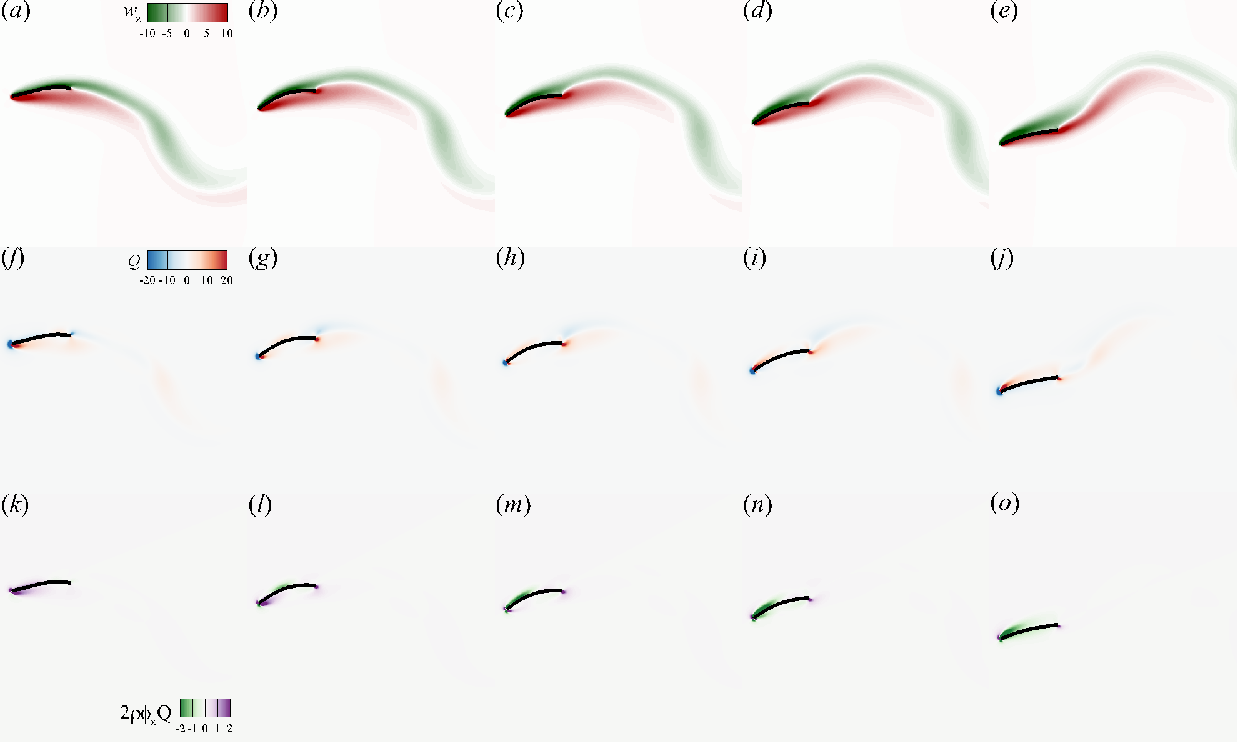}}
  \caption{Contours of ($a$-$e$) vorticity, ($f$-$j$) $Q$-criterion, and ($k$-$o$) $2\rho \phi_x Q$ for the membrane with $K_S=25$ under $St_c=0.2$ and $\theta_0=20^\circ$ at five time instants: ($a$,$f$,$k$)$t_1$, ($b$,$g$,$l$)$t_2$,($c$,$h$,$m$)$t_3$, ($d$,$i$,$n$)$t_4$, and ($e$,$j$,$o$)$t_5$ marked in figure~\ref{Fig7_CT_t}.
 }
\label{Fig11_St02_wQ}
\end{figure}

Figures~\ref{Fig11_St02_wQ} show the flow field snapshots at the five representative time instants marked in Figure~\ref{Fig7_CT_t}(d) for $St_c = 0.2$ with $\theta_0 = 20^\circ$ and $K_S = 25$. In contrast to the case of $St_c= 0.32$ shown in figures~\ref{Fig9_S25_wQ}, the vortex structures on both the upper and lower surfaces of the membrane become significantly weaker, resulting in much smaller Q-induced force, as shown in figures~\ref{Fig11_St02_wQ}($k$-$o$). On the other hand, the deformation of the membrane is also notably reduced, and the decrease of $St_c$ leads to a decrease in acceleration, which both result in a smaller body-acceleration force. As a result, as shown in figure~\ref{Fig7_CT_t}($d$), both $\widetilde{C}_{T,a}$ and $\widetilde{C}_{T,Q}$ exhibit very small amplitudes. Therefore, the time-averaged thrust coefficient increases monotonically with $St_c$, as illustrated in figure~\ref{Fig6_avgCT}($b$).

\subsection{Scaling laws for propulsive performance metrics}\label{scalinglaw}

\begin{figure}[htbp]
  \centerline{\includegraphics[width=0.8\textwidth]{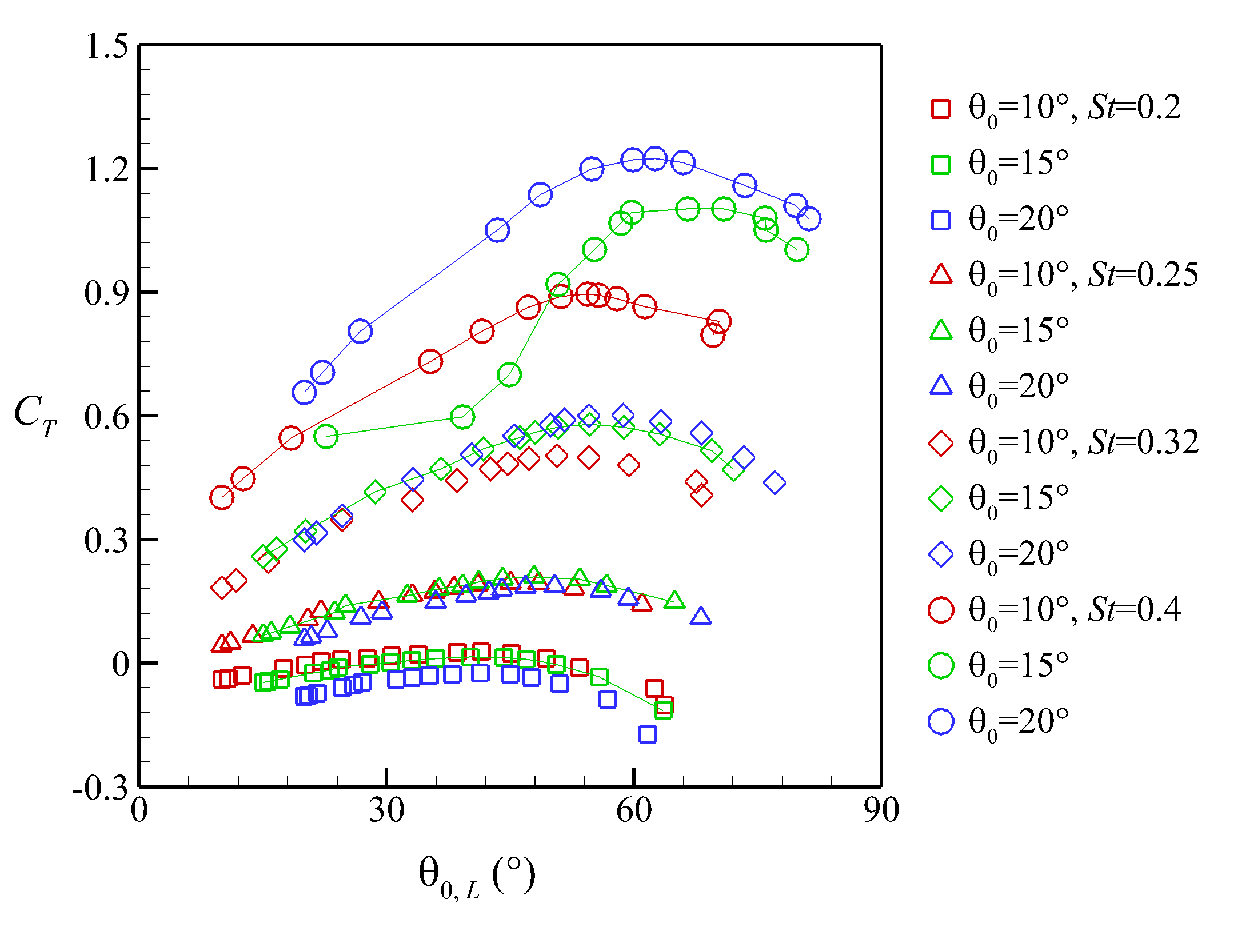}}
  \caption{Variation of $C_T$ with $\theta_{0,L}$.}
\label{Fig12_thetaLE_CT.eps}
\end{figure}

Due to the attachment of the vorticity layer near the front part of the membrane and the contours of $2\rho \phi_x Q$ indicating that the Q-induced force is mainly generated near the leading edge, the tangential angle at the leading edge is used to establish a correlation with the propulsive performance. Specifically, the effect of deformation is incorporated into the pitching amplitude $\theta_0$, resulting in a modified leading-edge tangential angular amplitude denoted as $\theta_{0,L}$. Figure~\ref{Fig12_thetaLE_CT.eps} shows the variation of the time-averaged thrust coefficient $C_T$ with respect to $\theta_{0,L}$ for four different $St_c$ values. For $St_c$ ranging from $0.2$ to $0.32$, the $C_T$–$\theta_{0,L}$ curves for three different pitching amplitudes collapse onto a single curve, and there exists an optimal $\theta_{0,L}$ that maximizes $C_T$. For $St_c = 0.4$, although the $C_T$–$\theta_{0,L}$ curves for different pitching amplitudes differ more noticeably, they still follow a similar trend and yield close optimal $\theta_{0,L}$ values. Based on these observations, $\theta_{0,L}$ can serve as a key parameter for constructing a scaling law for the force and power.

\cite{floryan2017scaling} established the scaling relationships between the thrust and power coefficients and the motion parameters for purely heaving or pitching rigid foils, by considering lift-based and added mass forces. Building on this framework, \cite{van2019scaling} developed scaling laws for rigid foils undergoing combined heaving and pitching motions. The corresponding expressions for the thrust and power coefficients are as follows:
\begin{equation}
\label{Eqn:scaling0CT}
C_T = c_1 St^2+c_2 St_h a_{\theta}^{*} \sin{\phi}+c_3 St_{\theta} a_{\theta}^{*}-c_4 a_{\theta}^{*},
\end{equation}
\begin{equation}
\label{Eqn:scaling0CP}
C_P = c_5 St^2+c_6 St_c St_h St_{\theta} \sin{\phi}+c_7 St_h a_{\theta}^{*} \sin{\phi}+c_8 St_c St_h^2+c_9 St_c St_{\theta}^2+c_{10} St_{\theta} a_{\theta}^{*},
\end{equation}
where $c_1$ to $c_{10}$ are the constants obtained from present numerical simulation results, $a_{\theta}^{*}$ is the pitching amplitude, $\phi$ is the phase difference between pitching and heaving motions, and the four Strouhal numbers are defined as: $St_c=f c/U_{\infty}$, $St_h=2 f h_0/U_{\infty}$, $St_{\theta}=2f c a_{\theta}^{*}/U_{\infty}$, $St=St_h^2+St_{\theta}+2 St_h St_{\theta} \cos{\phi}$. For the thrust equation~(\ref{Eqn:scaling0CT}), the first term is a simplified representation of the combined effects of part of the steady lift, the unsteady lift, foil acceleration, centrifugal force, and part of the Coriolis force. The second term arises from the steady lift, the third term corresponds to another part of the Coriolis force, and the fourth term accounts for viscous drag. The power equation (\ref{Eqn:scaling0CP}) is then derived based on the force expression and the kinematics of the foil motion. In the original scaling law for rigid foils, the non-dimensional angular amplitude $a_{\theta}^{*}$ corresponds to the prescribed pitching amplitude $\theta_0$. In the present scaling law for flexible membranes, $\theta_0$ is replaced by the amplitude of the leading-edge tangential angle $\theta_{0,L}$. This substitution accounts for the passive deformation of the membrane and provides a more representative measure of the effective pitching motion.

\begin{figure}[htbp]
    \centering
    \includegraphics[width=\textwidth]{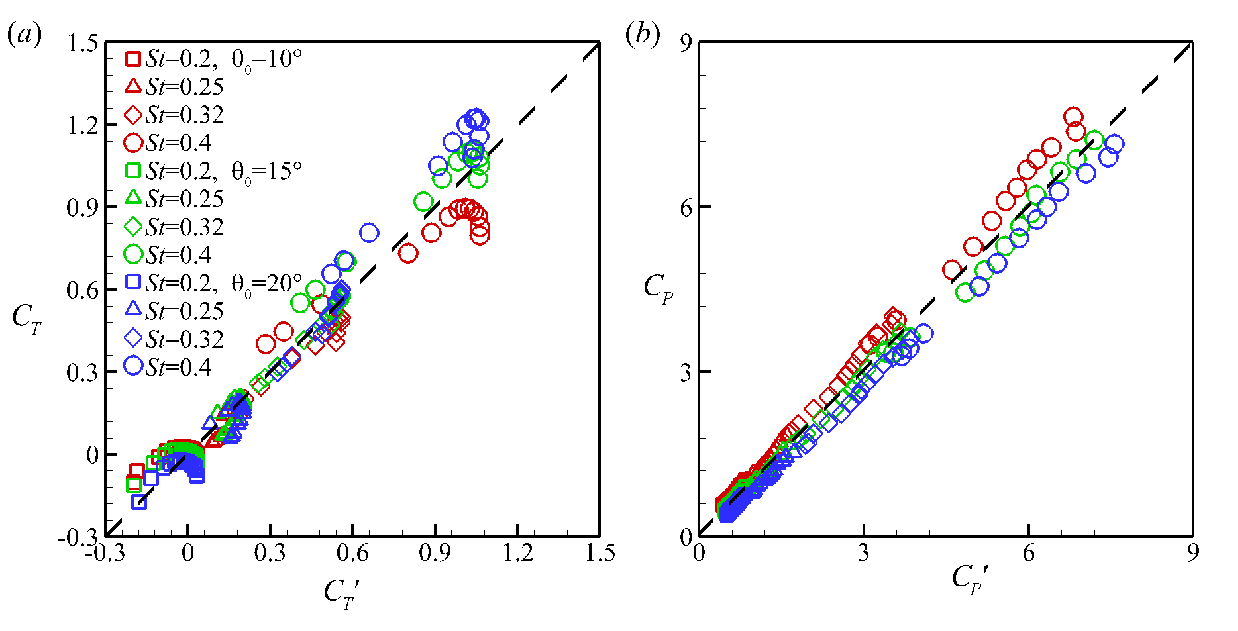}
    \caption{Scaling of the ($a$) thrust  and ($b$) power coefficients for all motion types.}
    \label{Fig10_scaling}
\end{figure}

Figure~\ref{Fig10_scaling}($a$) shows $C_T$ from the present numerical simulations versus $C_T'$ predicted by the scaling relation in Equation~(\ref{Eqn:scaling0CT}). The coefficients obtained from linear regression are $c_1 = 0.017902$, $c_2 = -7.570222$, $c_3 = -0.900580$, and $c_4 = 1.294311$. The small value of $c_1$ can be attributed to the choice of using the amplitude of the leading-edge tangential angle as the angular amplitude. Due to the bulging deformation of the membrane, the local angles between each membrane segment and the leading edge are generally smaller than this tangential angle. Since the first term in the scaling relation accounts for multiple force contributions—including lift, foil acceleration, and other inertial effects—this choice effectively amplifies their influence. As a result, the regression yields a small coefficient $c_1$ to compensate for the overestimated contribution of these effects. The two terms with the largest weights—corresponding to part of the steady lift and the viscous force—highlight the dominant thrust and drag contributions under the current scaling. After replacing the angular amplitude with the amplitude of the leading-edge tangential angle, the steady lift becomes the primary source of thrust, while the viscous drag serves as the main resistance. Figure~\ref{Fig10_scaling}($b$) shows $C_P$ from the present numerical simulations versus $C_P'$ predicted by the scaling relation in Equation~(\ref{Eqn:scaling0CP}). The coefficients obtained from linear regression are $c_5 = 2.017188$, $c_6 = -23.795996$, $c_7 = -0.311083$, $c_8 = 34.837700$. $c_9 = -2.049297$, and $c_{10} = -0.637621$. The data collapse closely along the $x = y$ reference line, indicating a good agreement between the scaling law and the numerical results. However, as $St_c$ increases to $0.4$, the agreement becomes less accurate compared to cases with smaller $St_c$. This discrepancy arises because, as shown in figure~\ref{Fig12_thetaLE_CT.eps}($d$), the $C_T$–$\theta_{0,L}$ curves for different pitching amplitudes at $St_c = 0.4$ do not collapse onto a single curve as they do for lower $St_c$ values. In summary, using the amplitude of the leading-edge tangential angle as an effective angular amplitude serves as a robust normalization approach. This substitution enables the collapse of data across a wide range of parameters, including pitching amplitude, Strouhal number, and membrane stretching stiffness. As a result, the proposed scaling laws can reliably predict the aerodynamic performance under diverse kinematic and structural conditions.

\section{Concluding remarks}\label{sec:Conclusion}
In summary, we numerically investigated the effects of the stretching coefficient (aeroelastic number) $K_S$, pitching amplitude $\theta_0$, and Strouhal number $St_c$ on the propulsive performance of a compliant membrane undergoing combined heaving and pitching oscillations in a uniform flow. The key findings are summarized as follows. 

For each combination of $\theta_0$ and $St_c$, there exist different optimal stretching stiffness coefficients that maximize the thrust coefficient $C_T$ and the propulsive efficiency $\eta$, respectively. The power coefficient $C_P$ increases monotonically with $K_S$, indicating that the enhanced thrust is achieved at the expense of greater power input from the membrane to the fluid. After introducing the frequency ratio $f^*$, defined as the ratio between the prescribed flapping frequency and the membrane's natural frequency, It was found that the maximum deflection across the chord over a flapping cycle can be approximately collapsed onto a single curve as a function of $f^*$, and increases monotonically with $f^*$. The instantaneous maximum deflection, calculated using the Young–Laplace equation with the pressure term approximated from thin-airfoil theory, shows good agreement with the numerical simulation results. However, the deviation increases with $St_c$, primarily because the analytical model assumes the membrane deformation to follow a symmetric arc shape, whereas the numerical simulations reveal increasingly asymmetric deformation profiles as $St_c$ increases. Furthermore, the optimal $f^*$ that maximizes the time-averaged thrust coefficient increases with $St_c$, whereas the optimal $f^*$ for maximizing the propulsive efficiency remains nearly constant around $f^* \approx 0.5$. Notably, all the optimal values of $f^*$ are below unity, indicating that the peak performance occurs in the pre-resonant regime.

Then, a force decomposition method based on a weighted integral of the second invariant of the velocity gradient tensor is adopted to decompose the membrane force into four components: Q-induced force, body-acceleration force, frictional force, and pressure viscous force. It is found that the Q-induced thrust coefficient $C_{T,Q}$ exhibits a similar non-monotonic trend with respect to $K_S$ as the total thrust coefficient $C_T$, it first increases and then decreases as $K_S$ decreases. This indicates that excessive deformation reduces the Q-induced thrust. Flow field snapshots show that the vorticity layer covers the upper surface of the membrane, while a negative-$Q$ region appears ahead of the leading edge. The part of this region that lies along the lower surface contributes positively to the Q-induced thrust. Together with the upper-surface vorticity layer, it generates a Q-induced thrust greater than that produced by the leading-edge vortex on the rigid plate. However, the bowl-like deformation of the membrane tends to trap the leading-edge vortex beneath the lower surface, thereby increasing the Q-induced drag. When $K_S$ is too small, this trapped vortex becomes even stronger, further enhancing the drag. As a result, a moderate value of $K_S$ maximizes the time-averaged Q-induced thrust coefficient. In addition, membrane stretching accelerates more fluid and contributes to a greater body-acceleration-induced thrust. Therefore, the total time-averaged thrust coefficient is also maximized at a moderate $K_S$. The optimal thrust occurs when the membrane deformation is large but not excessive, i.e., before resonance.

Finally, it is found that the thrust coefficients corresponding to different pitching amplitudes can be effectively collapsed onto a single curve when plotted against the leading-edge tangential angle, except at the highest Strouhal number ($St_c = 0.4$) considered in this study. Moreover, since the vortex layer remains mostly attached near the front portion of the membrane, these findings motivate the use of the leading-edge tangential angle as a key parameter in constructing scaling laws for both the thrust coefficient and the power coefficient. Therefore, in the scaling laws for thrust and power coefficients previously developed for rigid flapping plates, the leading-edge tangential angle is substituted for the angular amplitude. This substitution enables the collapse of data across a wide range of pitching amplitude, Strouhal number, and membrane stretching stiffness. As a result, the proposed scaling laws can reliably predict the aerodynamic performance of the membrane.

\begin{bmhead}[Acknowledgements.]
We thank Varghese Mathai from the University of Massachusetts Amherst for inspiring the direction. We gratefully acknowledge the financial support from the Max Planck Society and the German Research Foundation (DFG) through grants 521319293, 540422505, and 550262949. All the simulations have been conducted on the HPC systems of the Max Planck Computing and Data Facility (MPCDF). C. Zhang and A. Gao also acknowledge the support from the National Natural Science Foundation of China (Grants No.12102169 and No.12302320). 
\end{bmhead}

\begin{bmhead}[Declaration of Interests.]
The authors report no conflict of interest.  
\end{bmhead}

\bibliographystyle{jfm}
\bibliography{jfm}

\end{document}